\documentclass[aps,prl,twocolumn]{revtex4}
\usepackage{amsfonts}
\usepackage{amsmath}
\usepackage{graphicx}
\usepackage{dsfont}
\usepackage{diagbox}
\usepackage{array}
\usepackage{amssymb}
\usepackage{bbm}
\usepackage{float}
\usepackage{url}
\usepackage{hyperref}
\usepackage{booktabs}
\usepackage{color}
\usepackage{ulem}

\begin{document}

\title{Two-stage melting of an inter-component Potts long-range order in two dimensions}
\author{Feng-Feng Song$^{1}$ and Guang-Ming Zhang$^{1,2,3}$}
\email{gmzhang@tsinghua.edu.cn}
\affiliation{$^{1}$State Key Laboratory of Low-Dimensional Quantum Physics and Department
of Physics, Tsinghua University, Beijing 100084, China. \\
$^{2}$Collaborative Innovation Center of Quantum Matter, Beijing, China.\\
$^{3}$Frontier Science Center for Quantum Information, Beijing 100084, China.}
\date{\today }

\begin{abstract}
Interplay of topology and competing interactions can induce new phases and
phase transitions at finite temperatures. We consider a weakly coupled
two-dimensional hexatic-nematic XY model with a relative $Z_3$ Potts degrees
of freedom, and apply the matrix product state method to solve this model
rigorously. Since the partition function is expressed as a product of
two-legged one-dimensional transfer matrix operator, an entanglement entropy
of the eigenstate corresponding to the maximal eigenvalue of this transfer
operator can be used as a stringent criterion to determine various phase
transitions precisely. At low temperatures, the inter-component $Z_3$ Potts
long-range order (LRO) exists, indicating that the hexatic and nematic fields
are locked together and their respective vortices exhibit quasi-LRO. In the
hexatic regime, below the BKT transition of the hexatic vortices, the
inter-component $Z_3$ Potts LRO appears, accompanying with the binding of
nematic vortices. In the nematic regime, however, the inter-component $Z_3$
Potts LRO undergoes a two-stage melting process. An intermediate Potts
liquid phase emerges between the Potts ordered and disordered phases,
characterized by an algebraic correlation with formation of charge-neutral
pairs of both hexatic and nematic vortices. These two-stage phase
transitions are associated with the proliferation of the domain walls and
vortices of the relative $Z_3$ Potts variable, respectively. Our results
thus provide a prototype example of two-stage melting of a two-dimensional
long-range order, driven by multiple topological defects.
\end{abstract}

\maketitle

\textit{Introduction.} -The concept of topological defects has been one of
the cornerstones in the study of phase transitions in two-dimensional (2D)
systems. The known example is the Berezinskii-Kosterlitz-Thouless (BKT)
phase transition\cite{Berezinsky_1971,Kosterlitz_1973,Kosterlitz_1974} where
unbinding of integer vortices occurs in the absence of spontaneous breaking
of continuous symmetry. The followed renormalization group
Kosterlitz-Thouless-Halperin-Nelson-Young theory\cite%
{Halperin_1978,Nelson_1979,Young_1979} describes a two-stage melting process
of 2D crystals: dissociation of dislocation pairs induces a BKT-type
transition from a crystal phase to a hexatic phase with a quasi-long-range
(quasi-LRO) orientational order and a further transition is resulted from
unbinding of disclinations, as a dislocation can be represented as a coupled
pair of disclinations. Despite the four decade research, it is still
challenging to find a simple microscopic model that exhibit these phenomena
theoretically\cite{Kosterlitz_2016,Ryzhov_2017}.

\begin{figure}[tbp]
\centering
\includegraphics[width=0.45\textwidth]{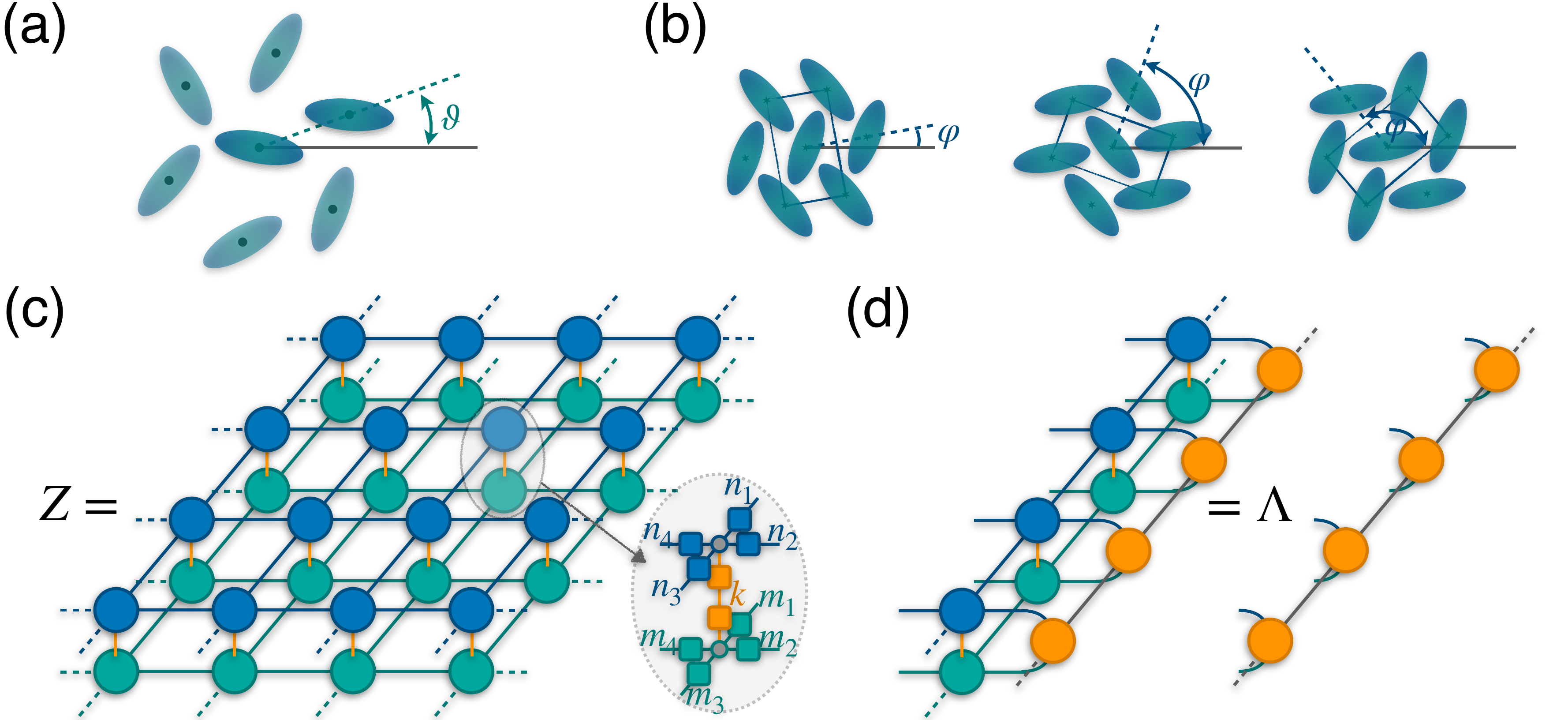}
\caption{(a) The bond-orientational angle $\protect\vartheta$ joining the
centers of neighboring molecules with respect to a laboratory axis. (b)
Three possible herringbone patterns. (c) The double-layer tensor network of
the partition function and the local tensor. (d) Eigen-equation of the 1D
transfer operator.}
\label{fig:tensor}
\end{figure}

Recently it has known that the interaction of multiple topological excitations
also plays a central role in many physical systems ranging from
superfluids/superconductors to condensed atom-molecular mixtures\cite{Strandburg_1988,
Bernard_2011,Chae_2012,Radzihovsky_2008,Shahbazi_2006,Parny_2016,Serna_2017,Kobayashi_2019}.
Among these studies, a prototype model is a 2D coupled hexatic-nematic
XY spin model proposed for an unusual melting process\cite{Bruinsma_1982,Aeppli_1984}
and a hidden-order phase transition of isotropic liquid-crystal thin films\cite%
{Jiang_1993,Jiang_1996,Touchette_2022}. As shown in Fig.\ref{fig:tensor}(a),
the hexatic degrees of freedom describe a six-fold bond-orientational field
represented by $\Theta=|\Theta|\exp(i6\vartheta)$ with $\vartheta$ as the bond
angle linking the centers of mass of neighboring molecules\cite{Halperin_1978},
while the nematic degrees of freedom arising from the herringbone order in the
crystalline phase\cite{Bruinsma_1982} are described by $\Phi=|\Phi|\exp(i2\varphi)$.
Three inequivalent herringbone patterns are displayed in Fig.\ref{fig:tensor}(b),
from which a different herringbone pattern with the same orientation can be
obtained by translating over a lattice vector. The Hamiltonian is described by a
coupled bilayer system
\begin{eqnarray}
H &=&-J_{2}\sum_{\langle i,j\rangle }\cos (2\varphi _{i}-2\varphi
_{j})-J_{6}\sum_{\langle i,j\rangle }\cos (6\vartheta _{i}-6\vartheta_{j})
\notag \\
&&-K\sum_{i}\cos (6\vartheta _{i}-6\varphi _{i}),  \label{eq:hn_model}
\end{eqnarray}
where $\varphi_{i}$ and $\vartheta_{i}\in \lbrack 0,2\pi ]$ are two $U(1)$
phase fields, $J_{2}$ and $J_{6}$ are their respective nearest-neighbour
intra-field couplings, and the inter-component coupling $K$ denotes the minimal
hexatic-nematic coupling allowed by the relative symmetry\cite{Bruinsma_1982}.

The study of this 2D coupled XY spin model in the context of nematic/hexatic
liquid crystalline systems has a very long history, and is plagued with complexities,
confusions and unexplained puzzling results, which have been reviewed in
Ref.\cite{Touchette_2022}. In the \textit{strongly} coupled model, large-scale Monte
Carlo simulations had determined the possible $Z_{3}$ Potts transition relative
to the BKT transition\cite{Jiang_1993,Jiang_1996}. In the hexatic regime, the
inter-component $Z_{3}$ Potts LRO is developed below the BKT transition of the hexatic
vortices, however, in the nematic regime the formation of the $Z_{3}$ Potts LRO coincides
with the BKT vortex binding transition\cite{Jiang_1993,Jiang_1996}. A recent numerical
work suggests that the $Z_{3}$ Potts LRO in the nematic regime may form above the BKT transition\cite{Touchette_2022}. Due to the lack of sharp thermodynamic signatures
associated with the binding of topological defects, the scenario of a single
transition cannot be fully excluded. The phase structure in the nematic regime
is expected to be revealed only in the \textit{weakly} coupled model, which reminds
largely unexplored because of the requirements of significantly larger system sizes
with enlarged vortex cores.

In order to resolve this long-standing issue, we apply a state-of-art tensor
network method\cite{Verstraete_2008,Orus_2014,Haegeman-Verstraete2017} to surmount
those difficulties. As the partition function of a 2D statistical model can be
represented as a product of two-legged 1D transfer operator (Fig.\ref{fig:tensor}(c)),
its eigen-equation (Fig.\ref{fig:tensor}(d)) can be solved by the algorithm
of variational uniform matrix product states (MPS) in the thermodynamic limit%
\cite{VUMPS,Fishman_2018,Laurens_Haegeman_2019,Laurens_Bram_2019}. According
to the singularity displayed by the entanglement entropy for the 1D quantum
analogue, the various phase transitions in the global phase diagram can be
precisely determined\cite{Li_2020,Song_2022_2}. In the hexatic regime, we obtain
the similar results to the previous ones\cite{Jiang_1993,Jiang_1996,Touchette_2022}.
To our surprise, in the nematic regime, we discover that the inter-component Potts
LRO itself undergoes a two-stage melting process. An intermediate Potts liquid
phase emerges simultaneously with the formation of charge-neutral pair of hexatic
and nematic vortices. Such two-stage transitions are associated with the separated
proliferations of the domain walls and vortices of the Potts variable, respectively.
Furthermore, we show that, when increasing the hexatic-nematic coupling strength, the
two-stage transitions gradually merge into a single transition, the result for the
strongly coupled model\cite{Jiang_1993,Jiang_1996}.

\textit{Tensor Network Method.} -We first simplify the model Hamiltonian
Eq.(1) by introducing new variables $\phi=2\varphi$ and $\theta=6\vartheta$,
\begin{eqnarray}
H/J&=&-\Delta\sum_{\langle
i,j\rangle}\cos(\phi_i-\phi_j)-(2-\Delta)\sum_{\langle
i,j\rangle}\cos(\theta_i-\theta_j)  \notag \\
&&-\lambda\sum_{i}\cos(\theta_i-3\phi_i),
\end{eqnarray}
where $\lambda=K/J$, $J=(J_2+J_6)/2$, and $\Delta=J_2/J$ as the relative
interacting strength of the hexatic and nematic fields. In the absence of
the hexatic-nematic coupling, the model has $U(1)\times U(1)$ symmetry and
exhibits two independent BKT phase transitions. For $\lambda\ne0$, the model
is invariant under a $U(1)\times Z_3$ transformation, $\phi_i\to
\phi_i+\alpha/3+{2\pi}k_i/3$ and $\theta_i\to\theta_i+\alpha$, where $k_i=0,
1, 2$ correspond to a $Z_3$ degrees of freedom. Specially, a sufficient
large $\lambda$ tends to lock the hexatic and nematic fields, $%
\theta_i=3\phi_i$, and the model is reduced to a generalized XY model\cite%
{Grannato_1986,Poderoso_2011,Canova_2014,Canova_2016}.

In the tensor network framework, after a duality transformation, the
partition function is expressed as a tensor contraction over all auxiliary
links,
\begin{equation}
Z=\mathrm{tTr}\prod_{i}O_{n_{1}m_{1},n_{2}m_{2}}^{n_{3}m_{3},n_{4}m_{4}}(i),
\label{eq:TN}
\end{equation}
where ``$\text{tTr}$'' denotes the tensor contraction, which forms a
double-layer tensor network shown in Fig.~\ref{fig:tensor}(c). Each local
tensor $O$ is given by
\begin{eqnarray}
O_{n_{1}m_{1},n_{2}m_{2}}^{n_{3}m_{3},n_{4}m_{4}} &=&\sum_{k}\left(
\prod_{l=1}^{4}I_{n_{l}}(\beta \Delta)I_{m_{l}}(\beta (2-\Delta))\right)
^{1/2}  \notag \\
&&\times I_{k}(\beta \lambda)\delta_{n_{1}+n_{2}}^{n_{3}+n_{4}+3k}\delta
_{m_{1}+m_{2}+k}^{m_{3}+m_{4}},
\end{eqnarray}
where $I_n(x)$ is the modified Bessel function of the first kind, $n$ and $m$
are integers, and $\beta=1/(k_B T)$. The global $U(1)\times Z_3$ invariance
is encoded in the local tensor.
In the thermodynamic limit, the partition function is determined by the
dominant eigenvalues of the 1D quantum transfer operator {$\hat{T}$}, whose
eigen-equation (Fig.~\ref{fig:tensor}(d)) {$\hat{T}|\Psi (A)\rangle =\Lambda
_{\max }|\Psi (A)\rangle$} can be accurately solved by the algorithm of
variational uniform MPS\cite%
{VUMPS,Fishman_2018,Laurens_Haegeman_2019,Laurens_Bram_2019}. The
corresponding 1D quantum Hamiltonian $\hat{H}_{1D}=-(1/\beta) \ln \hat{T}$,
and the leading eigenvector $|\Psi(A)\rangle$ is represented by a MPS, whose
precision is controlled by the auxiliary bond dimension $D$ of the local
tensors.

From the maximal eigenvalue and its eigenvector of the 1D quantum transfer
operator, various physical quantities can be estimated accurately\cite%
{Li_2020,Song_2021,Song_2022,Song_2022_2}. As the phase transitions are
concerned, the quantum entanglement entropy is the most efficient measure%
\cite{Vidal_2003,Pollmann_2009}, which can be directly determined via the
Schmidt decomposition: $S_{E}=-\sum_{\alpha=1}^{D}s_{\alpha }^{2}\ln
s_{\alpha }^{2}$ with $s_{\alpha}$ as the singular values from the
bipartition of the state $|\Psi(A)\rangle$. Various two-point correlation
functions can be evaluated by the trace of an infinite sequence of channel
operators containing two local impurity tensors\cite{SM}.

The superfluid response is described by the spin stiffness as the second
derivative of the free energy density {\ $f=-(1/N\beta)\ln Z$} with respect
to a twist $v$ along a reference direction: $\rho _{s}=\frac{\partial ^{2}f}{%
\partial ^{2}v}|_{v=0}$. The twist needs to be imposed in a way that
respects the joint $U(1)$ invariance of the hexatic and nematic fields as
\begin{equation}
(\phi _{i},\theta _{i})\rightarrow (\phi _{i}+\vec{v}\cdot \vec{r}%
_{i},\theta _{i}+3\vec{v}\cdot \vec{r}_{i}),
\end{equation}%
where $\vec{r}_{i}$ is the position vector for the lattice site $i$ and $%
\vec{v}$ is a constant vector to increase the phase difference across each
neighboring hexatic spins $3$ times larger than nematic spins. So the jump
of spin stiffness is altered from the BKT predictions when the unbinding of
the $\theta $ and $\phi $ vortices happens separately\cite{Hubscher_2013}.

\begin{figure}[tbp]
\centering
\includegraphics[width=0.45\textwidth]{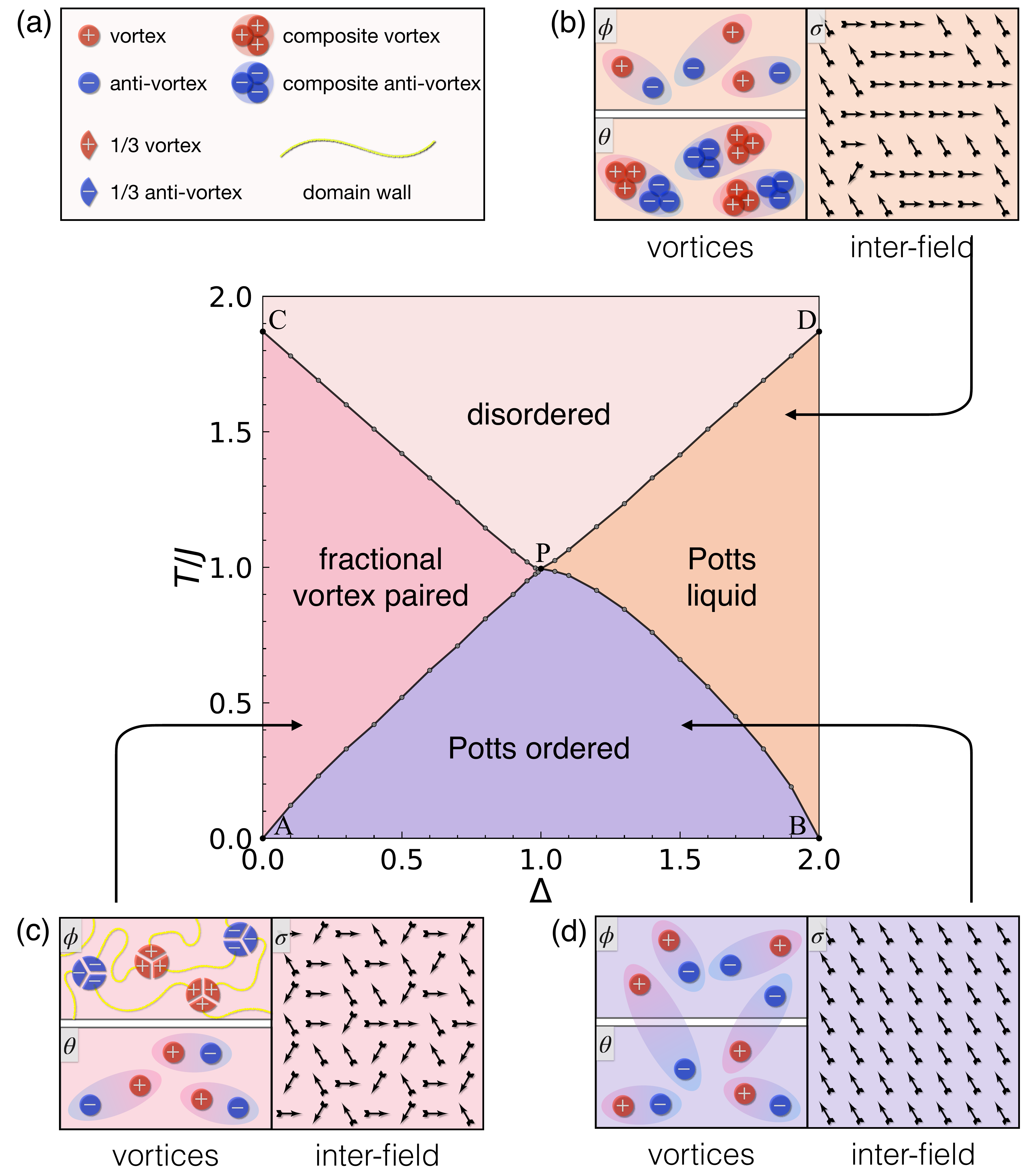}
\caption{The global phase diagram of the weakly coupled hexatic-nematic XY
model with a typical coupling $\protect\lambda=0.1$. (a) Schematic pictures
of different topological excitations. (b) In the Potts liquid phase, the
inter-component Potts variable has quasi-LRO, the vortex-antivortex pairs in
$\protect\phi$ fields are formed, and the dominant topological defects of
the $\protect\theta$ fields are composite vortex pairs with charge $q_{%
\protect\theta}=\pm3$. (c) In the fractional vortex paired phase, the $%
\protect\theta$ vortices have quasi-LRO, and the $\protect\phi$ vortices are
fractionalized as paired vortices with charge $q_{\protect\phi}=1/3$. (d) In
the Potts ordered phase, vortices in both types are bound in pairs,
accompanying the inter-component Potts LRO.}
\label{fig:phases}
\end{figure}

The hexatic-nematic XY model has a rich physics, and the most intriguing
results are expected in the weakly coupled case. For a typical value of $%
\lambda =0.1$, a global phase diagram is derived in Fig.~\ref{fig:phases}.
All phase boundaries are determined by the singularity displayed in the
entanglement entropy $S_{E}$ for the 1D quantum transfer operator. Since the
succession of phases crucially depends on the intra-component coupling
ratio, our results are discussed in hexatic $\Delta <1$ and nematic $\Delta
>1$ regimes, separately.

\begin{figure}[tbp]
\centering
\includegraphics[width=0.45\textwidth]{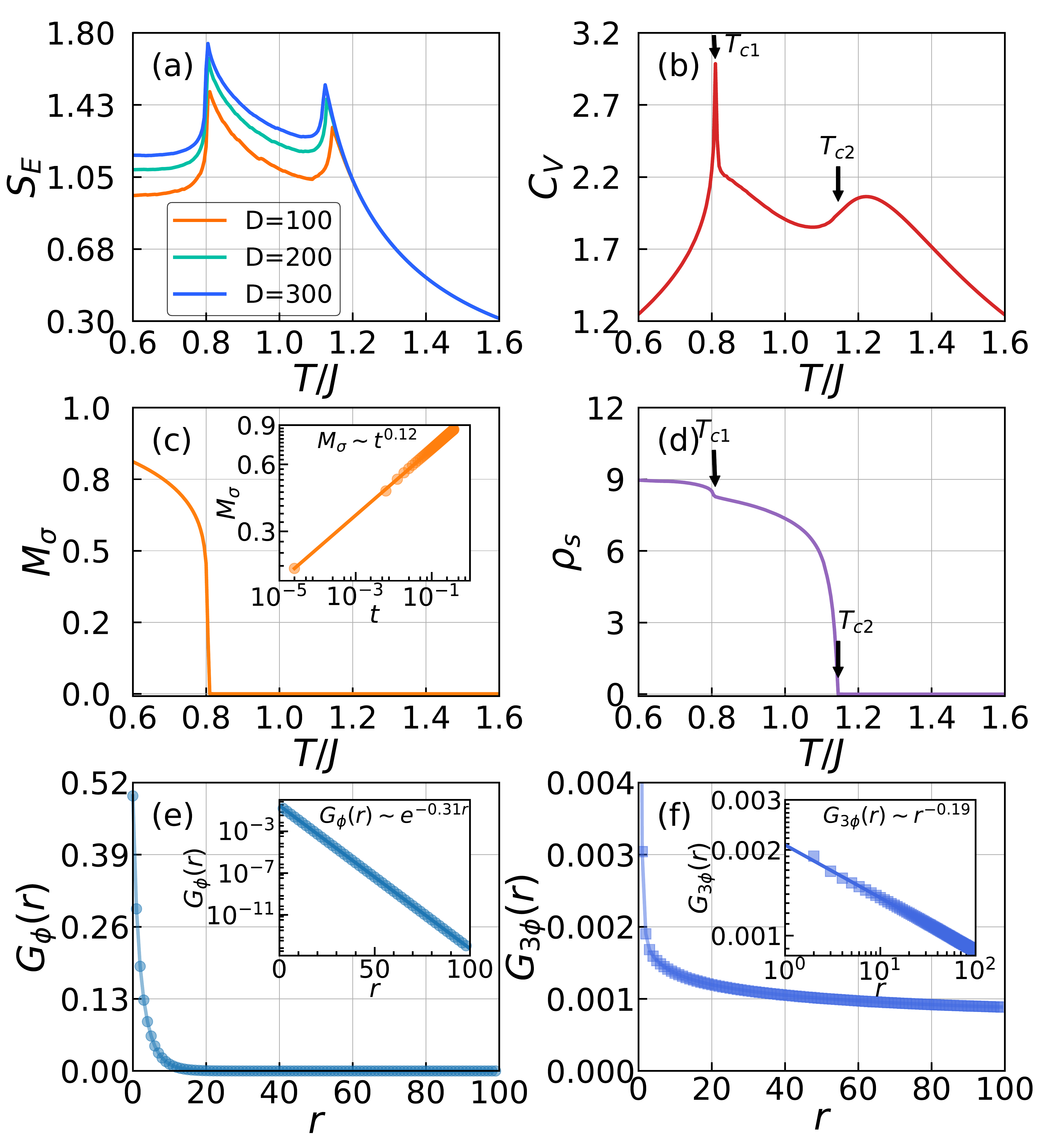}
\caption{ The numerical results for $\Delta=0.8$ and $\protect\lambda=0.1$.
(a) The entanglement entropy. (b) The specific heat. (c) The magnetization
of the Potts variable. {\ In the inset $t$ is the reduced temperature $%
t=|T_{c1}-T|/T_{c1}$.} (d) The superfluid stiffness. (e) and (f) At $T=1.0J$%
, the correlation function $G_{\protect\phi}(r)$ for $q_{\protect\phi}=1$
vortices decays exponentially, but the correlation function $G_{3\protect\phi%
}(r)$ for $q_{\protect\phi}=1/3$ vortices decays in power law.}
\label{fig:smDelta}
\end{figure}

\textit{Phase transitions in the hexatic regime.} -For a typical value $%
\Delta =0.8$, our numerical results show two singular peaks in the
entanglement entropy at $T_{c1}\simeq 0.805J$ and $T_{c2}\simeq 1.145J$,
respectively, as shown in Fig.~\ref{fig:smDelta}(a). The peak positions are
nearly unchanged with increasing bond dimensions $D=100,200,300$ so the
transition points are determined with high accuracy. The corresponding
specific heat is shown in Fig.~\ref{fig:smDelta}(b), exhibiting a sharp
divergence at $T_{c1}$ and a small rounded bump around $T_{c2}$. The maximum
of the hump is above $T_{c2}$, a typical feature of the BKT transition\cite%
{Kosterlitz_1973,Kosterlitz_1974}, while the low-$T$ singular behavior is
fitted by $C_{V}\propto|T-T_{c1}|^{\alpha }$ with $\alpha\simeq 1/3$.
Further evidence is provided by a $Z_{3}$ order parameter $M_{\sigma
}=\langle \cos (\sigma_{i})\rangle $ with $\sigma _{i}=\theta _{i}/3-\phi
_{i}$. As shown in Fig.~\ref{fig:smDelta}(c), $M_{\sigma }$ becomes finite
at $T_{c1}$, suggesting that a true LRO is established and the relative
phase between the $\phi$ and $\theta$ fields is fully locked. The
magnetization satisfies the scaling form {$M_{\sigma }\propto
(T_{c1}-T)^{\beta}$} with the critical exponent {$\beta \simeq 1/9$}. These
critical exponents are perfectly in agreement with the $Z_{3}$ Potts
transition\cite{Wu_1982}.

As shown in Fig.~\ref{fig:smDelta}(d), the spin stiffness starts to
dramatically increase from zero at the BKT transition $T_{c2}$. When the
temperature further decreases, a small jump appears to enhance the spin
stiffness at the Potts transition $T_{c1}$ precisely. Such a small increase
of spin stiffness is resulted from bindings of integer $\phi$ vortices\cite%
{Canova_2014,Touchette_2022}. When cooling the system from the disordered
phase, the correlation function $G_{\theta }(r)=\langle\cos
(\theta_{i}-\theta_{i+r})\rangle $ exhibits a power law decay at $T_{c2}$,
indicating the binding of hexatic vortices. In the intermediate temperature
regime ($T_{c1}<T<T_{c2}$), we found that $G_{\phi}(r)=\langle\cos
(\phi_{i}-\phi_{i+r})\rangle$ decays exponentially, but $G_{3\phi}(r)$
exhibits an algebraic correlation. In Fig.~\ref{fig:smDelta}(e) and (f), a
direct comparison at $T=1.0J$ suggests that the vortices are fractionalized
into the $q_{\phi}=1/3$ vortices. As the correlation length $\xi _{\phi }$
extracted from $G_{\phi}(r)$ follows an exponentially divergence above $%
T_{c1}$, we thus conclude that the low-temperature transition is a hybrid
BKT and Potts transition\cite{SM}.

\begin{figure}[t]
\centering
\includegraphics[width=0.45\textwidth]{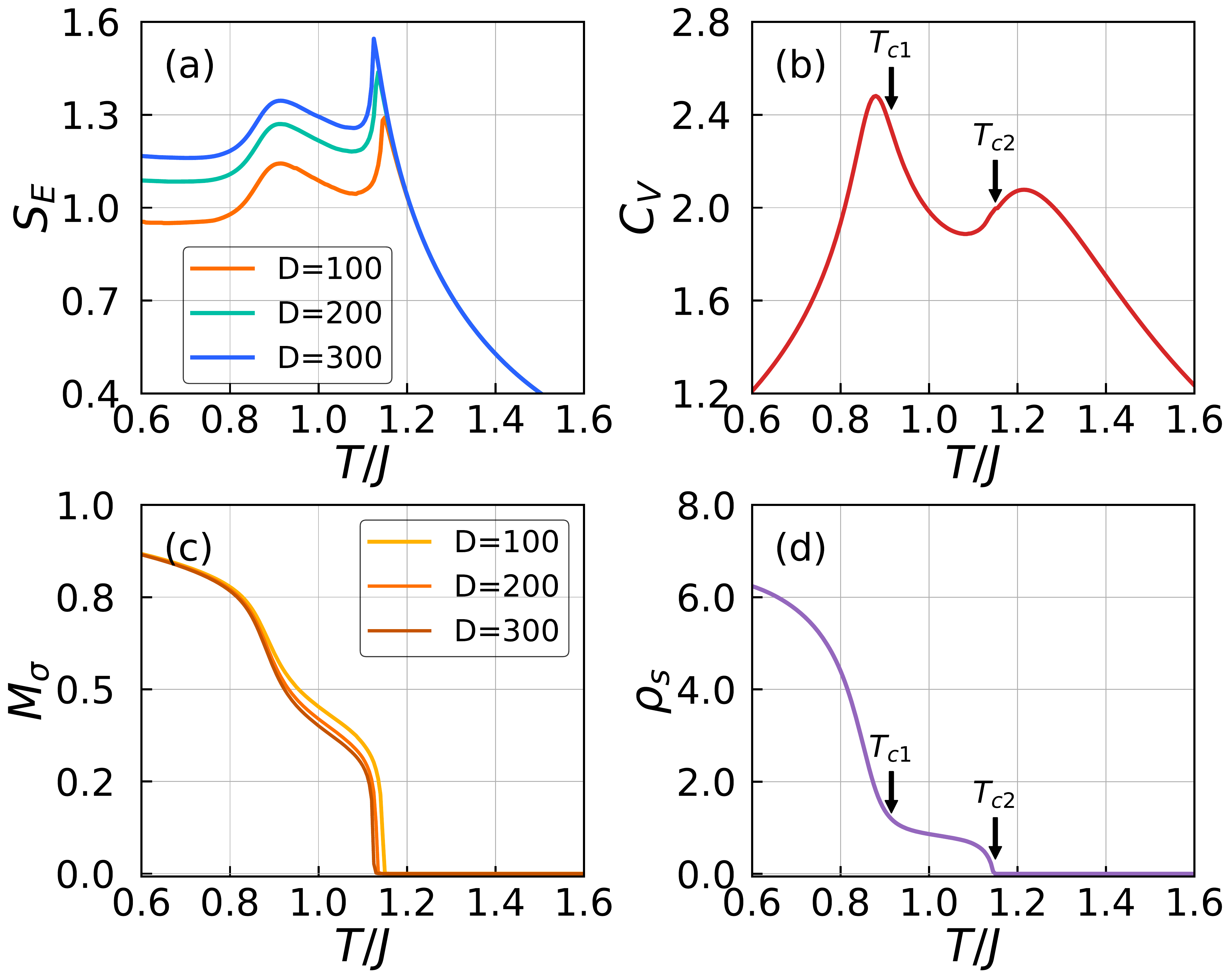}
\caption{ The numerical results for $\Delta=1.2$ with $\protect\lambda=0.1$.
(a) The entanglement entropy. (b) The specific heat. (c) The Potts
magnetization. (d) The superfluid stiffness. }
\label{fig:lgDelta}
\end{figure}

\textit{Phase transitions in the nematic regime.} -For a typical value $%
\Delta =1.2$, our numerical calculations show two peaks in the entanglement
entropy at $T_{c1}\simeq 0.915J$ and $T_{c2}\simeq 1.15J$, respectively, as
seen in Fig.~\ref{fig:lgDelta}(a). Unlike the strongly coupled model\cite%
{Touchette_2022}, the specific heat $C_{V}$ exhibits a pronounced peak at $%
T_{c1}$ and a bump at $T_{c2}$, as displayed in Fig.~\ref{fig:lgDelta}(b).
The bump appears lightly above $T_{c2}$ as the usual BKT transition, but the
peak locates below $T_{c1}$. Surprisingly, the magnetization of the relative
Potts variable displays a two-step feature below $T_{c2}$, as shown in Fig.~%
\ref{fig:lgDelta}(c). The artificial finite Potts magnetization in the intermediate
phase is due to the finite-bond dimension effect, which should decrease slowly 
to zero with increasing bond dimensions in MPS approximations for $U(1)$ phases\cite
{Laurens_Bram_2019,Li_2020}. To figure out the transition at $T_{c1}$, we
calculate the superfluid stiffness, which also has a two-step jump, as
displayed in Fig.~\ref{fig:lgDelta}(d). In contrast to the hexatic regime,
the stiffness drops more dramatically at $T_{c1}$ than at $T_{c2}$. Since
the phase twist $\vec{v}$ applied to the hexatic $\theta $ field is three
times larger than that to the nematic $\phi $ field, the first drop of the
stiffness is nearly six times larger than the second drop.

\begin{figure}[tbh]
\centering
\includegraphics[width=0.45\textwidth]{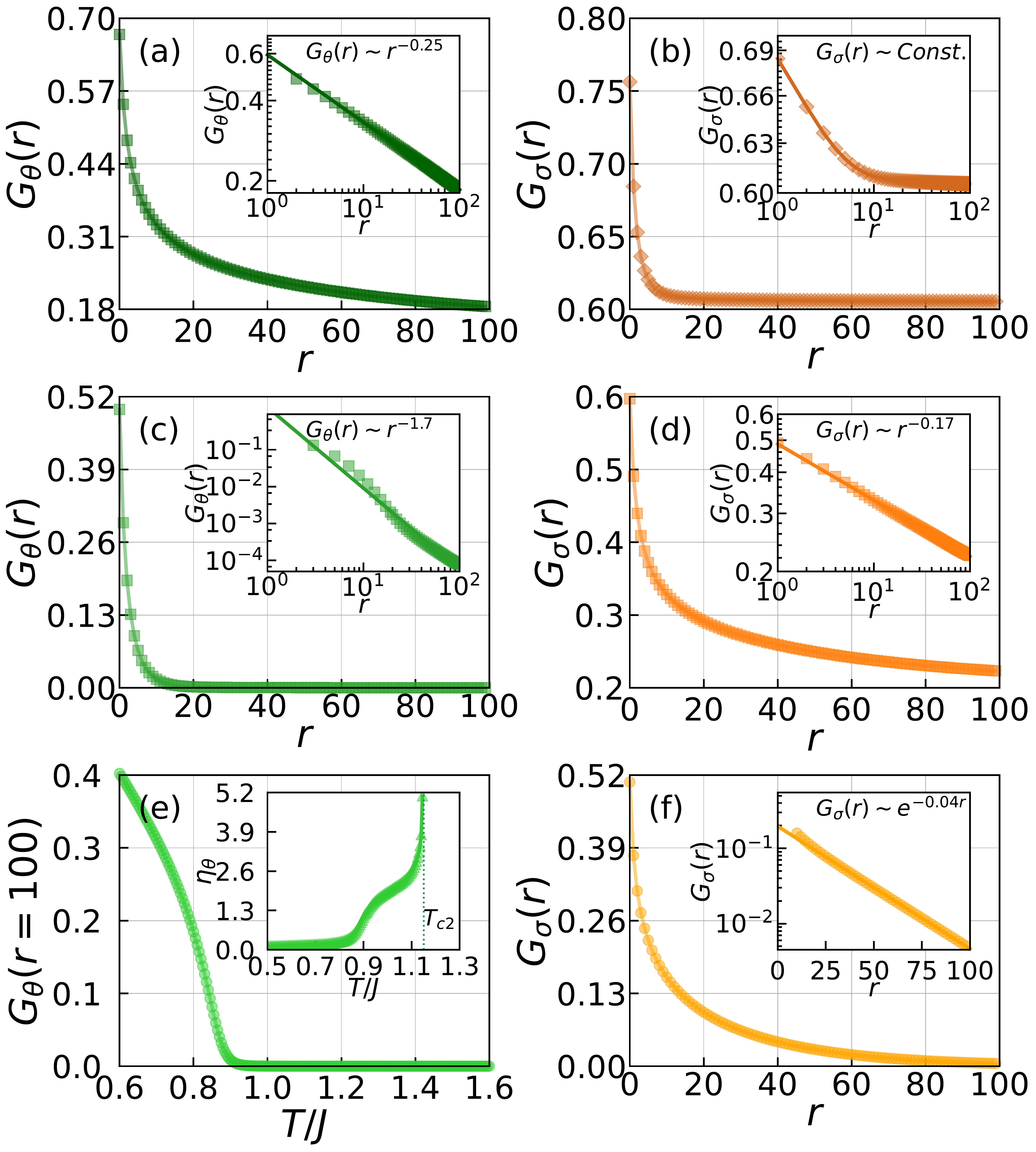}
\caption{ (a) and (c) The correlation functions $G_{\protect\theta}(r)$ in
both the Potts ordered phase ($T=0.8J$) and the Potts liquid phase ($T=1.0J$%
). (e) The correlation $G_{\protect\theta}(r)$ at $r=100$ and the exponents
in the inset. (b), (d) and (f) The correlation functions $G_{\protect\sigma%
}(r)$ in the Potts ordered, Potts liquid, and disordered phases.}
\label{fig:clfn}
\end{figure}

The abrupt drop of spin stiffness at $T_{c1}$ is rather rare, and the
binding of charge-neutral vortex pairs of $\theta$ field can be ruled out
from the correlation function $G_{\theta}(r)=\langle\cos(\theta_i-%
\theta_{i+r})\rangle$. On two sides of $T_{c1}$, the $G_{\theta}(r)$
correlations have the power law behavior as shown in Fig.~\ref{fig:clfn}(a)
and (c), indicating that the quasi-LRO persists through the transition. As
displayed in Fig.~\ref{fig:clfn}(e), a direct comparison between the
amplitudes of $G_{\theta}(r)\sim r^{-\eta_\theta}$ at $r=100$ indicates that
the correlation above $T_{c1}$ is greatly suppressed by three orders of
magnitude, and the exponent $\eta_{\theta}$ varies with temperature depicted
in the inset. The extremely weak correlation above $T_{c1}$ may account for
the dramatic jump in total stiffness at $T_{c1}$. Below $T_{c2}$, the onset
of algebraic correlations of the $\phi$ vortices also induces the quasi-LRO
of the $\theta $ vortices. So the dominant topological excitations between $%
T_{c1}$ and $T_{c2}$ are composite vortex pairs of $q_{\theta}=3$ with an
internal structure of three bound vortices.

By calculating the correlations of the relative $Z_3$ Potts variable, we
find that the inter-component Potts order-disorder phase transition splits
into two different transitions separated by an intermediate liquid phase
with quasi-LRO correlation. As shown in Fig.~\ref{fig:clfn}(b), (d) and (f),
the correlation function $G_{\sigma}(r)$ exhibits a LRO in the Potts ordered
phase, an algebraic decay in the intermediate phase, and an exponential
decay in the Potts disordered phase, respectively. Actually the Potts
quasi-LRO occurs simultaneously with the formation of charge-neutral pairs
of hexatic and nematic vortices. Moreover, the correlation length for Potts
variables displays a similar structure as the $p$-state clock model with $p>4
$ (Ref.\cite{SM}).

\begin{figure}[thbp]
\includegraphics[width=0.48\textwidth]{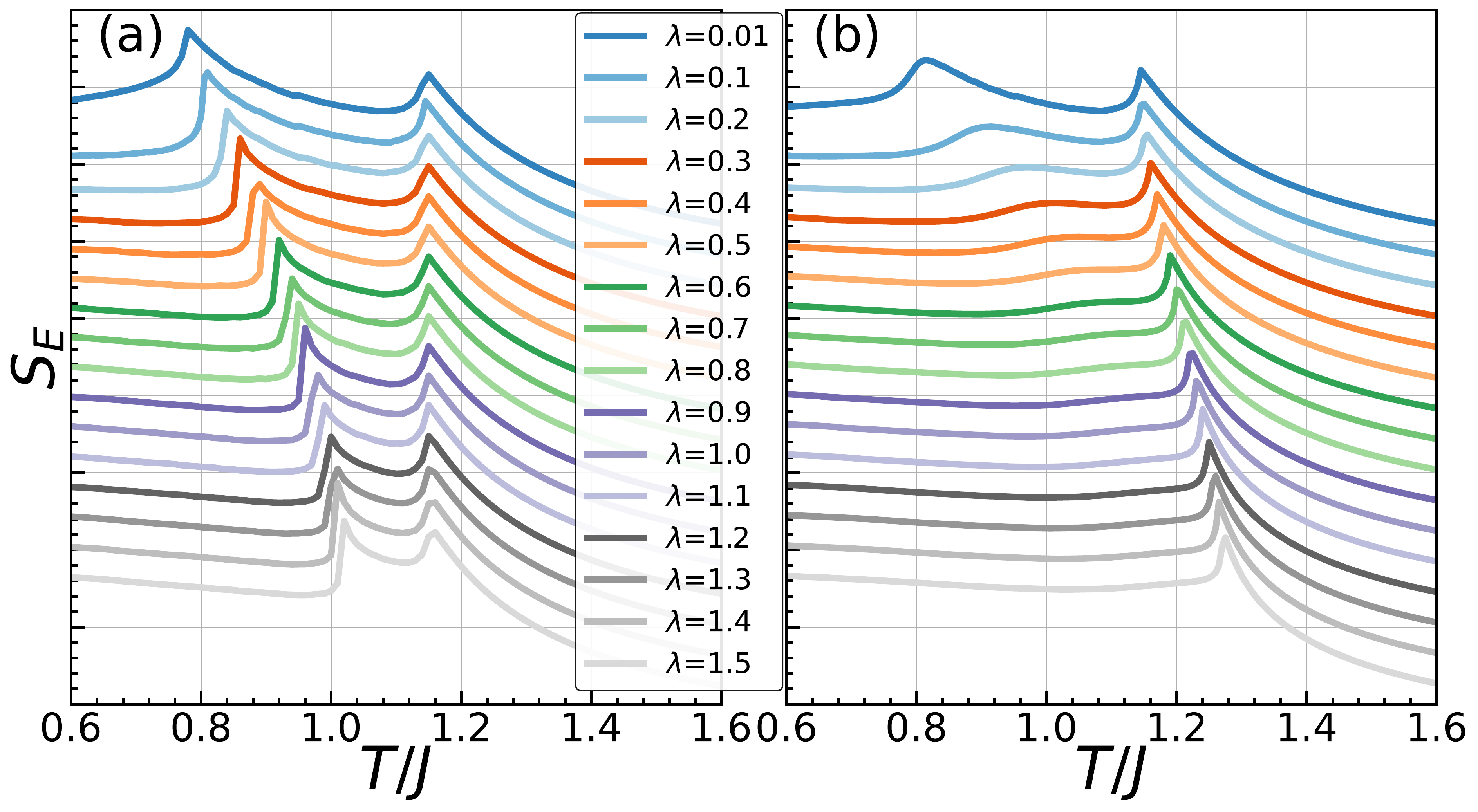}
\caption{The entanglement entropy for different $\protect\lambda$ in the
hexatic regime (a) $\Delta=0.8$ and the nematic regime (b) $\Delta=1.2$. For
a good comparison, each curve is relatively shifted by $\Delta {S_E}=0.2$.
The displayed results are obtained under MPS bond dimension $D=100$.}
\label{fig:lambda}
\end{figure}

Unlike the 2D Ising model, the excitations in the $Z_{3}$ Potts model
include both the loop-like domain walls and vortices, and the usual Potts
transition is resulted from a coupled proliferation of two topological
defects\cite{Einhorn-1980}. However, a quasi-LRO phase appears in a
generalized $Z_{3}$ Potts model by artificially raising the vortex core
energy\cite{Bhattacharya_2016}. We notice that the effective core energy of $%
Z_{3}$ vortices in the coupled hexatic-nematic model is related to the
hexatic-nematic coupling. A larger $\lambda $ makes the excitations of free
vortices in the $\theta$ field more costly, and the domain wall energy
increases dramatically to forbid the pre-excitations of domain walls.
However, a small $\lambda $ is required to achieve a relatively high ratio
of the core energy of the $Z_{3}$ vortices. To illustrate this physics, we
calculated the entanglement entropy as a function of temperature for
different values of $\lambda$. As shown in Fig.~\ref{fig:lambda}, in the
hexatic regime ($\Delta =0.8$), the entanglement entropy results indicate
that the the model always has two phase transitions similar to those of
strongly coupled model, while in the nematic regime ($\Delta=1.2$),
two-stage phase transitions are allowed only for a small $\lambda $ value.
As $\lambda $ increases, the intermediate Potts liquid phase obscures
because the lower transition gets vague, and two transitions are merged
together.

\textit{Conclusion and Outlook.} -We have studied a weakly coupled
hexatic-nematic XY model by using the tensor-network method. It is found
that, the hexatic regime shares the similar phase structure to that of the
strongly coupled model\cite{Jiang_1993,Jiang_1996,Touchette_2022}. In the
nematic regime, however, an inter-component $Z_{3}$ Potts liquid phase
emerges as the intermediate phase of the two-stage melting of the Potts LRO.
Actually this emergent Potts liquid phase opens up a promising route towards
better understanding the hidden structure of the phase transitions driven by
different topological defects. Such an intriguing splitting of phase
transitions should exist in other systems with multiple topological
excitations, such as multi-stage melting process of 2D liquid crystal films%
\cite{Pindak_1981,Huang_1981,Chou_1998}. Above the melting of positional
order, the orientational order itself possesses a further structure. Future
study can be extended to the coupled $U(1)\times Z_n$ systems with larger $n$
where a richer phase diagram is expected\cite{Nelson_1980}. Moreover, in
various XY systems like condensed atom-molecular mixtures and
multi-component superfluids/superconductors\cite%
{Donley_2002,Chin_2010,Fernandes_2010,Fernandes_2019}, such unconventional
phenomena should also be realizable.

\textbf{Acknowledgments.}
The research is supported by the National Key Research and Development Program of MOST of China (2017YFA0302902).

\widetext

\section{Appendix A: Tensor network representation of our model}

A good wealth of fascinating physics have been found in many-body systems but the numerical treatment still remains challenging due to the nature of strong correlations. Tensor networks (TN) have proven to be a very efficient  tool to overcome these challenges based on the variational ansatz wave functions. Due to the unique advantages of faithfully capturing the entanglement structure of many-body states, they are increasingly becoming a standard tool for not only strongly correlated quantum systems but also many-body classical problems.

The central but often nontrivial task in condensed matter physics is to evaluate the partition function of a system. To implement the TN method in the coupled hexatic-nematic XY model, the first step is to convert the classical lattice model with nearest-neighbor local interactions into a TN representation. The coupled hexatic-nematic XY model on a 2D square lattice is defined by the Hamiltonian
\begin{equation}
H =-J_{2}\sum_{\langle i,j\rangle }\cos (2\varphi _{i}-2\varphi _{j}) - J_{6}\sum_{\langle i,j\rangle }\cos (6\vartheta _{i}-6\vartheta_{j}) - K\sum_{i}\cos (6\vartheta _{i}-6\varphi _{i}),
\label{eq:Hamiltonian_1}
\end{equation}
where $\langle i,j\rangle$ refers to nearest neighbors, $\varphi _{i}$ and $\vartheta _{i}\in \lbrack 0,2\pi \rbrack$ are phase angles associated to the lattice site $i$, $J_{2}$ and $J_{6}$ are their respective intra-component coupling strength, and $K$ denotes the inter-component coupling. To lowest order in $\varphi$ and $\vartheta$ by introducing $\phi=2\varphi$ and $\theta=6\vartheta$, the simplified Hamiltonian in the same universality class can be written as
\begin{equation}
H/J=-\sum_{\langle i,j\rangle}\Delta\cos(\phi_i-\phi_j)-(2-\Delta)\sum_{\langle i,j\rangle}\cos(\theta_i-\theta_j)-\lambda\sum_{i}\cos(\theta_i-3\phi_i),
\label{eq:Hamiltonian_2}
\end{equation}
where $J=\frac{1}{2}(J_2+J_6)$, $\lambda=K/J$ and $\Delta=J_2/J\in[0, 2]$ tuning the relative strength of hexatic and nematic interactions.

\begin{figure}[tbp]
\centering
\includegraphics[width=.99\linewidth]{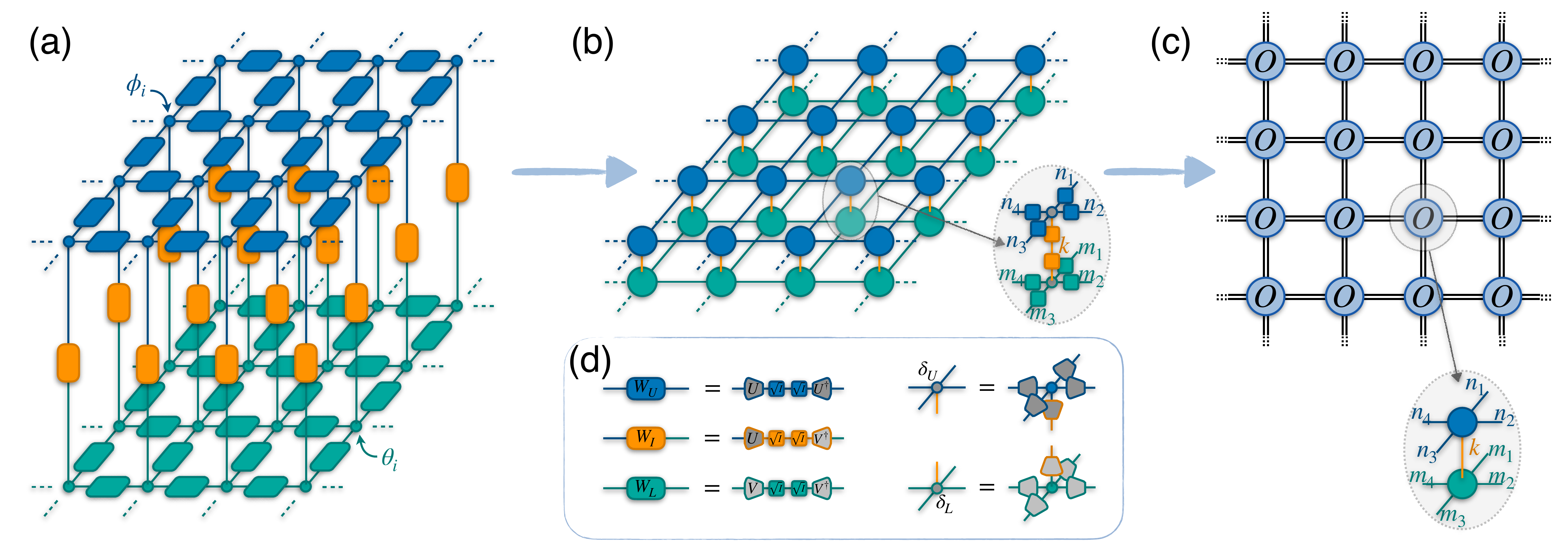}
\caption{
(a) Tensor network representation of the partition function with Boltzmann weights represented as a tensor product of local bonds.
(b) The double layer TN derived in the main text.
(c) The 2D uniform tensor network representation of the partition function.
(d) The eigenvalue decompositions of the symmetric matrix $W$ and the transformation to discrete degrees of freedom.
}
\label{fig:partition_function}
\end{figure}

To derive the TN representation, we express the partition function on its original lattice and the Boltzmann weights are represented as a tensor product of local bonds as shown in Fig.~\ref{fig:partition_function}(a)
\begin{equation}
Z=\iint \frac{\mathrm{d} \phi_{i} \mathrm{d} \theta_i}{(2 \pi)^2} \,\prod_{\langle i,j\rangle}W_U(\phi_i,\phi_j)W_I(\phi_i,\theta_i)W_L(\theta_i,\theta_j),
\end{equation}
where
\begin{align}
W_U(\phi_i,\phi_j) &= \mathrm{e}^{\beta \Delta \cos(\phi_i-\phi_j)},\\
W_I(\phi_i,\theta_i) &= \mathrm{e}^{\beta \lambda \cos(\theta_i-3\phi_i)},\\
W_L(\theta_i,\theta_j) &= \mathrm{e}^{\beta (2-\Delta) \cos(\theta_i-\theta_j)}
\end{align}
can be viewed as infinite matrices whose indices are the continuous $\phi$ and $\theta$ variables. The partition function is now cast into a double layer TN representation, where the integrations of $\int d\phi/2\pi$ and $\int d\theta_i/2\pi$ are denoted as blue and green dots and the matrix indices take the same values at the joint points.

The duality transformation changes the local tensors from continuous $U(1)$ variables onto discrete basis for further numerical analysis. As shown in Fig.~\ref{fig:partition_function}(d), we perform the character expansion for the  symmetric matrix $W$
\begin{align}
W_U(\phi_i,\phi_j) &= \sum_n U_{\phi_i,n}\,I_{n}(\beta \Delta)\,U_{\phi_j,n}^*,\\
W_I(\phi_i,\theta_i) &= \sum_k U_{\phi_i,-3k}\,I_{k}(\beta \lambda)\,V_{\theta_i,k}^*,\\
W_L(\theta_i,\theta_j) &= \sum_m V_{\theta_i,m}\,I_{m}(\beta (2-\Delta))\,V_{\theta_j,m}^*
\end{align}
where $U_{\phi_i,n}=\mathrm{e}^{in\phi_i}$, $V_{\theta_i,m}=\mathrm{e}^{im\theta_i}$, and the diagonal $I_n(x), I_m(x), I_k(x)$ are the modified Bessel functions of the first kind.

Then, as displayed in Fig.~\ref{fig:partition_function}(d), we can simply integrate out the $U(1)$ phase variables at each site by a Fourier transformation
\begin{align}
\int \frac{d \phi_i}{2\pi} U_{\phi_i,n_1}U_{\phi_i,n_2}U_{\phi_i,n_3}^*U_{\phi_i,n_4}^*U_{\phi_i,-3k}&= \delta_{n_1+n_2}^{n_3+n_4+3k}\equiv\delta_U,\\
\int \frac{d \theta_i}{2\pi} V_{\theta_i,m_1}V_{\theta_i,m_2}V_{\theta_i,m_3}^*V_{\theta_i,m_4}^*V_{\theta_i,k}&= \delta_{m_1+m_2+k}^{m_3+m_4}\equiv\delta_L.
\end{align}
In this way, the continuous variables $\phi$ and $\theta$ are now transformed into the discrete bond indices $n,m$ and $k$. Moreover, we evenly divide the diagonal $I$ tensors and take a contraction of the $\sqrt{I}$ tensors connecting to the respective $\delta$ tensors. As a result, the double layer TN representation of the partition function in the main text is obtained as shown in Fig.~\ref{fig:partition_function}(b). The local tensor $O$  can then be obtained by grouping the corresponding tensor indices in the upper and lower layers and summing out the inter-layer $k$ indices
\begin{equation}
O_{n_{1}m_{1},n_{2}m_{2}}^{n_{3}m_{3},n_{4}m_{4}}=\sum_{k}\left(\prod_{l=1}^{4}I_{n_{l}}(\beta \Delta)I_{m_{l}}(\beta (2-\Delta))\right) ^{1/2} I_{k}(\beta \lambda)\,\delta _{n_{1}+n_{2}}^{n_{3}+n_{4}+3k}\delta_{m_{1}+m_{2}+k}^{m_{3}+m_{4}}.
\end{equation}
And finally the partition function is now successfully converted into a uniform TN on the 2D square lattice as displayed in Fig.~\ref{fig:partition_function}(c)
\begin{equation}
Z=\mathrm{tTr}\prod_{i}O_{n_{1}m_{1},n_{2}m_{2}}^{n_{3}m_{3},n_{4}m_{4}}(i)
\end{equation}
where ``tTr'' denotes the sum over all auxiliary bond indices. The symmetries of the original model is well preserved in the TN representation. It is evident that the global $U(1)\times Z_3$ invariance of the bilayer model is encoded in each local tensor since $O_{n_{1}m_{1},n_{2}m_{2}}^{n_{3}m_{3},n_{4}m_{4}}\ne0$ only if $n_1 +3m_1 +n_2 +3m_2 = n_3 +3m_3 +n_4 +3m_4$.

As a matter of fact, the equivalence of the Hamiltonian \eqref{eq:Hamiltonian_1} and \eqref{eq:Hamiltonian_2} can be easily verified from the perspective of local tensors. The duality transformation can be performed on $\varphi$ and $\vartheta$ in the same way for Hamiltonian \eqref{eq:Hamiltonian_1}, and the integration will generate the same local tensor as
\begin{align}
\int \frac{d \varphi_i}{2\pi} \mathrm{e}^{i2n_1\varphi_i}\mathrm{e}^{i2n_2\varphi_i}\mathrm{e}^{-i2n_3\varphi_i}\mathrm{e}^{-i2n_4\varphi_i}\mathrm{e}^{-i6k\varphi_i} &= \delta_{2n_1+2n_2}^{2n_3+2n_4+6k} = \delta_{n_1+n_2}^{n_3+n_4+3k}\equiv\delta_U,\\
\int \frac{d \vartheta_i}{2\pi} \mathrm{e}^{i6m_1\vartheta_i}\mathrm{e}^{i6m_2\vartheta_i}\mathrm{e}^{-i6m_3\vartheta_i}\mathrm{e}^{-i6m_4\vartheta_i}\mathrm{e}^{-i6k\vartheta_i} &= \delta_{6m_1+6m_2+6k}^{6m_3+6m_4} = \delta_{m_1+m_2+k}^{m_3+m_4}\equiv\delta_L.
\end{align}
Therefore all the physical quantities deducted from the partition function should be the same for both \eqref{eq:Hamiltonian_1} and \eqref{eq:Hamiltonian_2}.

\section{Appendix B: Algorithm of uniform matrix product states}

Although it is often straightforward to write down the TN representation for the many-body problem, the real challenge for numerical simulations comes from the contraction of the whole tensor network which is proved to be NP hard when the network is complicated. Fortunately, a lot of algorithms have been proposed to contract an infinite translation-invariant TN. One of the best practices to address the challenge is the algorithm of uniform matrix product states\cite{VUMPS,Fishman_2018,Laurens_Haegeman_2019}, which greatly speeds up the process of finding the leading eigenvector of the transfer matrix based on a variational ansatz.

\begin{figure}[thbp]
\centering
\includegraphics[width=.88\linewidth]{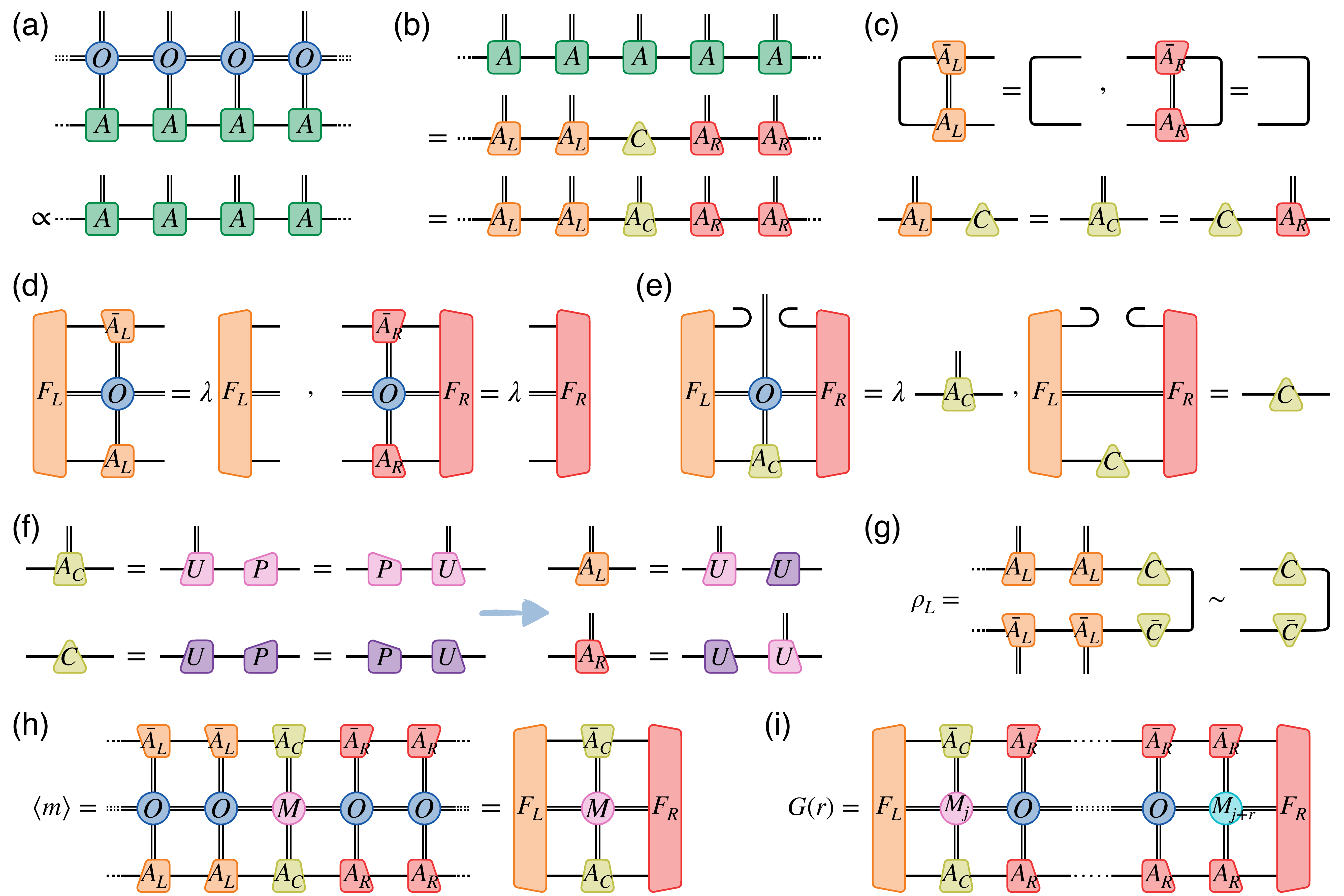}
\caption{
(a) Eigen-equation for the fixed-point uMPS of the transfer operator $T$.
(b) The uniform representation and two equivalent mixed canonical forms of the fixed-point uMPS.
(c) The canonical conditions of the fixed-point local tensors and the isometric gauge transformation between them.
(d) Eigen-equations to update the left and right environmental eigenvectors of the channel operators.
(e) Eigen-equations to obtain the central tensors based on the new environment.
(f) Polar decompositions to get the left and right canonical local tensors from the central tensors.
(g) The reduced density matrix obtained by tracing out the right part of the uMPS.
(h) Expectation value of a local observable by contracting the leading eigenvectors of the channel operators.
(i) Two-point correlation function expressed by contracting a sequence of channel operators. }
\label{fig:VUMPS}
\end{figure}

Here we provide a brief review of this efficient algorithm based on a set of optimized eigen-solvers. In the thermodynamic limit, the partition function is determined by the dominant eigenvalues of the 1D quantum transfer operator $\hat{T}$, whose eigen-equation is shown in Fig.~\ref{fig:VUMPS}(a)
\begin{equation}
\hat{T}|\Psi (A)\rangle =\Lambda _{\max }|\Psi (A)\rangle.
\end{equation}
For a translation-invariant system, the leading eigenvector of the transfer matrix can be expressed as the uniform matrix product states (uMPS):
\begin{equation}
|\Psi(A)\rangle=\sum_{\{(n_i m_i)\}}\mathrm{Tr}\left(\cdots A^{(n_1m_1)} A^{(n_2m_2)} A^{(n_3m_3)} \cdots\right),
\end{equation}
which is constructed by infinitely many repetitions of rank-3 $A_{\alpha\beta}^{(n_im_i)}$ with auxiliary bond dimension $\alpha,\beta=1,2,\ldots, D$. Here, $D$ is the upper bond dimension which controls the accuracy of the approximations. As shown in Fig.~\ref{fig:VUMPS}(b), to fix the gauge of the uMPS, we should bring the uMPS into the equivalent mixed canonical forms as
\begin{equation}
|\Psi\rangle=\sum_{\alpha,\beta=1}^{D}C_{\alpha,\beta}|\Psi_{\alpha}^{[-\infty,(n_jm_j)]}\rangle\otimes|\Psi_{\beta}^{[(n_{j+1}m_{j+1}),+\infty]}\rangle,
\end{equation}
and
\begin{equation}
|\Psi\rangle=\sum_{\alpha,\beta=1}^{D}\left(A_{C}^{(n_jm_j)}\right)_{\alpha,\beta}|\Psi_{\alpha}^{[-\infty,(n_{j-1}m_{j-1})]}\rangle\otimes|\Psi_{\beta}^{[(n_{j+1}m_{j+1}),+\infty]}\rangle,
\end{equation}
where $|\Psi_{\alpha}^{[-\infty,(n_jm_j)]}\rangle$ and $|\Psi_{\beta}^{[(n_{j+1}m_{j+1}),+\infty]}\rangle$ are the left and right orthonormal basis comprised of the left and right canonical local tensors
\begin{align}
|\Psi_{\alpha}^{[-\infty,(n_jm_j)]}\rangle&=\sum_{\{(n_i m_i)\}}\mathrm{Tr}\left(\cdots A_L^{(n_{j-1}m_{j-1})} A_L^{(n_jm_j)} \right),\\
|\Psi_{\beta}^{[(n_{j+1}m_{j+1}),+\infty]}\rangle&=\sum_{\{(n_i m_i)\}}\mathrm{Tr}\left(A_R^{(n_{j+1}m_{j+1})} A_R^{(n_{j+2}m_{j+2})} \cdots\right).
\end{align}

The left and right orthonormal tensors $A_L$ and $A_R$ satisfy the isometric constraints as displayed in Fig.~\ref{fig:VUMPS}(c)
\begin{align}
&\sum_{(nm)}\left(A_L^{(nm)}\right)^\dagger A_L^{(nm)}=\mathcal{I},\\
&\sum_{(nm)}A_R^{(nm)}\left(A_R^{(nm)}\right)^\dagger=\mathcal{I}.
\end{align}
Another fixed point equation is obtained by pulling the central tensor from right to left
\begin{equation}
A_C^{(nm)}=A_L^{(nm)}C=C A_R^{(nm)}.
\end{equation}

The VUMPS algorithm contains three key steps summarized in Fig.~\ref{fig:VUMPS}(d)-(f) to find the fixed-point tensors $A_L$, $A_R$ and $C$ of the transfer operator $\hat{T}$. These three steps (d)-(f) are repeated sequentially  until the error is less than a given convergence threshold.

(1). As shown in Fig.~\ref{fig:VUMPS}(d), we calculate the left and right environments $F_L$ and $F_R$ by solving the eigenvalue equation with the Arnoldi method:
\begin{align}
\mathbb{T}_L F_L &= \lambda F_L,\\
\mathbb{T}_R F_R &= \lambda F_R,
\end{align}
where $\mathbb{T}_L$ and $\mathbb{T}_R$ are the channel operators of $A_L$ and $A_R$.

(2). As shown in Fig.~\ref{fig:VUMPS}(e), we calculate the central tensors $A_C$ and $C$ based on the new environment by the Arnoldi method:
\begin{align}
H_{A_C} A_C &= \lambda A_C,\\
H_C C &= C,
\end{align}
where $H_{A_C}$ and $H_C$ are the effective environment comprised of the updated $F_L$ and $F_R$.

(3). As shown in Fig.~\ref{fig:VUMPS}(f), we use the left and right polar decompositions of new $A_C$ and $C$ to update $A_L$ and $A_R$ satisfying the isometric conditions
\begin{align}
A_C &= U_{A_C}^{[l]}P_{A_C}^{[l]}=P_{A_C}^{[r]}U_{A_C}^{[r]},\\
C &= U_{C}^{[l]}P_{C}^{[l]}=P_{C}^{[r]}U_{C}^{[r]},
\end{align}
where $U_{A_C}^{[l]}$,  $U_{A_C}^{[r]}$, $U_{C}^{[l]}$ and $U_{C}^{[r]}$ are unitary matrices, and the matrices $P_{A_C}^{[l]}$,  $P_{A_C}^{[r]}$, $P_{C}^{[l]}$ and $P_{C}^{[r]}$ are hermitian and positive.
Finally, we obtain
\begin{align}
A_L &=U_{A_C}^{[l]}U_{C}^{[l]\dagger},\\
A_R &=U_{C}^{[r]\dagger}U_{A_C}^{[r]}.
\end{align}

Actually, the row-to-row transfer matrix should play the same role as the matrix product operator for the 1D quantum spin chains whose logarithmic form can be mapped to a 1D
quantum system with complicated interactions. With such a correspondence, the finite-temperature properties of the 2D statistical problem are related to a 1D quantum model at zero temperature and all the low-temperature physics can be achieved from the fixed-point uMPS.

\section{Appendix C: Calculations of physical quantities}

Once the fixed-point uMPS is achieved, various physical quantities can be precisely calculated because the 2D network can be easily squeezed into a 1D chain of channel operators based on the fixed uMPS.

By mapping the transfer matrix to a 1D quantum transfer operator, we bring to it modern concepts of quantum entanglement, while also taking advantage of the sharp criterion for detecting various phase transitions. As shown in Fig.~\ref{fig:VUMPS}(g), we perform a bipartition on the MPS and trace out the right part using the right canonical condition. Since the $|\Psi^{[-\infty,(n_jm_j)]}\rangle$ also forms a unitary basis, the reduced density matrix $\rho_L$ of the left part is expressed as
\begin{equation}
\rho_L = \mathrm{Tr}_R |\Psi\rangle\langle \Psi|=|\Psi^{[-\infty,(n_jm_j)]}\rangle C C^\dagger \langle\Psi^{[-\infty,(n_jm_j)]}|\sim C C^\dagger.
\end{equation}
And the entanglement entropy\cite{Vidal_2003} is readily obtained by
\begin{equation}
S_E=-\mathrm{Tr}\,\left(\rho_L\ln\rho_L\right)=-\sum_{\alpha=1}^{D}s_{\alpha}^2\ln s_{\alpha}^2,
\end{equation}
where $s_\alpha$ are the singular values of the $C$ matrix.

With the fixed-point uMPS, the contraction of an infinite 2D tensor network can be reduced to the trace of an infinite 1D chain of channel operators. The expectation value of of a single-site observable $m(\phi_j,\theta_j)$
\begin{equation}
\langle m(\phi_j,\theta_j)\rangle=\frac{1}{Z}\prod_{i}\iint \frac{\mathrm{d}\phi _{i}\mathrm{d}\theta_{i}}{\left( 2\pi \right)^{2}}\prod_{\langle i,j\rangle }\mathrm{e}^{\beta \Delta\cos (\phi _{i}-\phi _{j})}  \mathrm{e}^{\beta (2-\Delta)\cos (\theta _{i}-\theta_{j})}\mathrm{e}^{\beta \lambda\cos (\theta _{i}-3\phi _{i})}m(\phi_j,\theta_j)
\end{equation}
can be obtained by the contraction of a reduced network containing an impurity tensor thanks to the fixed point environment tensors $F_L$ and $F_R$ as displayed in Fig.~\ref{fig:VUMPS}(h)
\begin{equation}
\langle m\rangle=\mathrm{Tr}\left(\cdots\mathbb{T}_{L}\mathbb{T}_{L}\mathbb{T}_{M}\mathbb{T}_{R}\mathbb{T}_{R}\cdots\right)=\langle F_L|\mathbb{T}_{M}|F_R\rangle
\end{equation}
where $\mathbb{T}_{M}=\mathrm{tTr}(A_C\otimes M\otimes \bar{A}_C)$. And the corresponding impurity tensor $M$ can be achieved by
\begin{equation}
M_{n_1m_1, n_2m_2}^{n_3m_3, n_4m_4}=
\sum_k\left(\prod_{l=1}^{4} I_{n_l}(\beta \Delta)I_{m_l}(\beta (2-\Delta))\right)^{1/2}I_{k}(\beta \lambda)
\iint \frac{\mathrm{d} \phi \mathrm{d} \theta}{(2 \pi)^2}\,
\mathrm{e}^{i\phi(n_1+n_2-n_3-n_4-3k)}
\mathrm{e}^{i\theta(m_1+m_2+k-m_3-m_4)}m(\phi,\theta).
\end{equation}
For instance, the corresponding impurity tensors for $m_{\phi}=\mathrm{e}^{i\phi}$, $m_{\theta}=\mathrm{e}^{i\theta}$ and $m_{\sigma}=\mathrm{e}^{i(\theta/3-\phi)}$ are
\begin{align}
\left(M_\phi\right)_{n_{1}m_{1},n_{2}m_{2}}^{n_{3}m_{3},n_{4}m_{4}}&=\sum_{k}\left(\prod_{l=1}^{4}I_{n_{l}}(\beta \Delta)I_{m_{l}}(\beta (2-\Delta))\right) ^{1/2} I_{k}(\beta \lambda)\,\delta _{n_{1}+n_{2}+1}^{n_{3}+n_{4}+3k}\delta_{m_{1}+m_{2}+k}^{m_{3}+m_{4}},\\
\left(M_\theta\right)_{n_{1}m_{1},n_{2}m_{2}}^{n_{3}m_{3},n_{4}m_{4}}&=\sum_{k}\left(\prod_{l=1}^{4}I_{n_{l}}(\beta \Delta)I_{m_{l}}(\beta (2-\Delta))\right) ^{1/2} I_{k}(\beta \lambda)\,\delta _{n_{1}+n_{2}}^{n_{3}+n_{4}+3k}\delta_{m_{1}+m_{2}+k+1}^{m_{3}+m_{4}},\\
\left(M_\sigma\right)_{n_{1}m_{1},n_{2}m_{2}}^{n_{3}m_{3},n_{4}m_{4}}&=\sum_{k}\left(\prod_{l=1}^{4}I_{n_{l}}(\beta \Delta)I_{m_{l}}(\beta (2-\Delta))\right) ^{1/2} I_{k}(\beta \lambda)\,\delta _{n_{1}+n_{2}}^{n_{3}+n_{4}+3k+1}\frac{\sqrt{3}}{2\pi}\frac{(-1)^l}{l+1/3},
\end{align}
where $l=m_1+m_2+k-m_3-m_4$.

Moreover, the two-point correlation function $G(r)=\langle m_j m_{j+r}\rangle$ between local observables $m_j(\phi_j, \theta_j)$ and $m_{j+r}(\phi_{j+r}, \theta_{j+r})$
\begin{equation}
G(r)=\frac{1}{Z}\prod_{i}\iint \frac{\mathrm{d}\phi _{i}\mathrm{d}\theta_{i}}{\left( 2\pi \right)^{2}}\prod_{\langle i,j\rangle }\mathrm{e}^{\beta \Delta\cos (\phi _{i}-\phi _{j})}  \mathrm{e}^{\beta (2-\Delta)\cos (\theta _{i}-\theta_{j})}\mathrm{e}^{\beta \lambda\cos (\theta _{i}-3\phi _{i})}m_j(\phi_j,\theta_j)m_{j+r}(\phi_{j+r},\theta_{j+r}).
\end{equation}
can be calculated in the same way as a contraction of a 1D chain with two local impurity tensors $M_j$ and $M_{j+r}$ as shown in Fig.~\ref{fig:VUMPS}(i)
\begin{equation}
G(r)=\langle F_L|\mathbb{T}_{M_j}\underbrace{\mathbb{T}_{R}\cdots \mathbb{T}_{R}}_{r-1}\mathbb{T}_{M_{j+r}}|F_R\rangle
\end{equation}
where $\mathbb{T}_{M_j}=\mathrm{tTr}(A_C\otimes M_j\otimes \bar{A}_C)$ and $\mathbb{T}_{M_{j+r}}=\mathrm{tTr}(A_R\otimes M_{j+r}\otimes \bar{A}_R)$.

\section{Appendix D: Particular calculation of spin stiffness}

Although there is no local order parameter due to the absence of the $U(1)$ symmetry breaking, the onset of the superfluidity of the hexatic-nematic XY model can be described by the spin stiffness $\rho_s$\cite{Fisher_1973, Nelson_1977}, which characterizes the change of free energy in response to a small uniform phase twist $v$ along a reference direction
\begin{equation}
\rho_s=\left.\frac{\partial^2 f_v}{\partial v^2}\right\vert_{v=0} = -\frac{1}{N\beta }\left[\frac{1}{Z_v}\frac{\partial^2 Z_v}{\partial v^2}-\left(\frac{1}{Z_v}\frac{\partial Z_v}{\partial v}\right)^2\right]_{v=0}
\label{eq:stiffness}
\end{equation}
where $f_v=-\frac{1}{N\beta}\ln Z_v$ is the free energy per site.

Since the two components of $\phi$ and $\theta$ are coupled together via a relevant inter-component term, the contribution of the total spin stiffness of both fields should be taken into account. In order to preserve the $U(1)\times Z_3$ invariance of the model, the phase twist should be applied as
\begin{equation}
(\phi_i,\theta_i)\to(\phi_i+\vec{v}\cdot\vec{r}_i, \theta_i+3\vec{v}\cdot\vec{r}_i)
\end{equation}
where $\vec{r}_i$ is the position vector for the lattice site $i$ and $\vec{v}$ is a constant vector to increase the phase difference across each neighboring hexatic spins $3$ times larger than nematic spins. When a global twist is applied along $y$-axis, the phase differences between two nearest-neighbor sites in $y$-direction are increased by $v$ for $\phi$ field and $3v$ for $\theta$ field, and the corresponding Hamiltonian is changed as
\begin{align}
\begin{split}
H_v=&-\Delta\sum_{\langle i,j\rangle_x}\cos(\phi_i-\phi_j)-(2-\Delta)\sum_{\langle i,j\rangle_x}\cos(\theta_i-\theta_j)-\lambda\sum_{i}\cos(\theta_i-3\varphi_i)\\
&-\Delta\sum_{\langle i,j\rangle_y}\cos(\phi_i-\phi_j+v)-(2-\Delta)\sum_{\langle i,j\rangle_x}\cos(\theta_i-\theta_j+3v).
\end{split}
\end{align}
In the same time, the TN representation of the partition function is modified as
\begin{equation}
Z_v=\mathrm{tTr}\prod_{i}(O_v)_{n_{1}m_{1},n_{2}m_{2}}^{n_{3}m_{3},n_{4}m_{4}}(i),
\end{equation}
where an additional phase factor is appended on the vertical leg of the local tensor
\begin{equation}
(O_v)_{n_{1}m_{1},n_{2}m_{2}}^{n_{3}m_{3},n_{4}m_{4}} =\mathrm{e}^{i(n_1+3m_1)v}O_{n_{1}m_{1},n_{2}m_{2}}^{n_{3}m_{3},n_{4}m_{4}}.
\end{equation}

\begin{figure}[thbp]
\centering
\includegraphics[width=.88\linewidth]{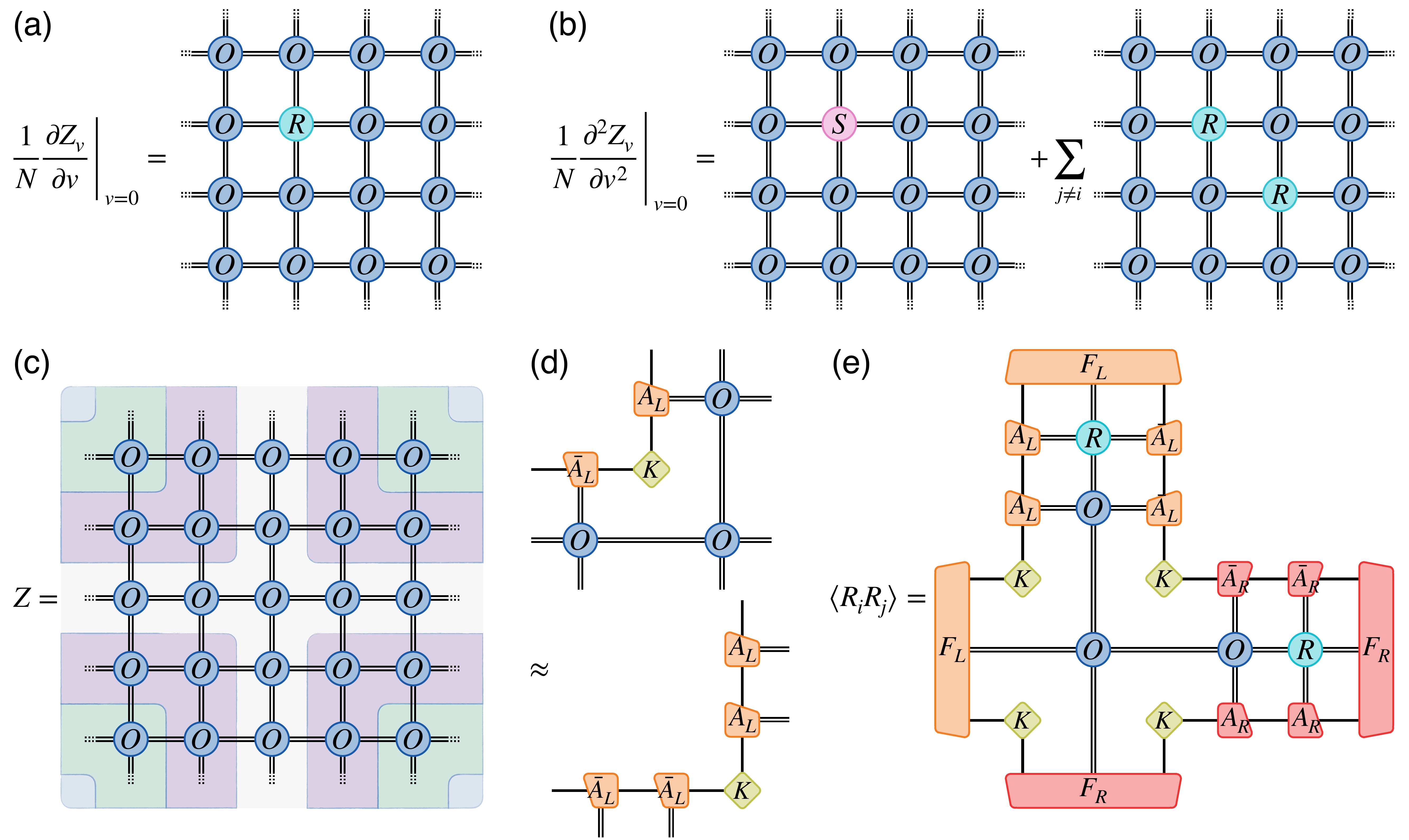}
\caption{
(a) The tensor network representation of the first derivative of the partition function.
(b) The tensor network representation of the second derivative of the partition function.
(c) The contraction of an infinite tensor network using corner environments from four corners.
(d) The eigen-equation of the top left corner-shaped fixed point.
(e) The computation of a two-point function expressed by contracting the channel operators and the corner tensors.}
\label{fig:stiffness}
\end{figure}

The precision of the direct second differentiation on the partition function often suffers from a finite differentiation step $v$. Alternatively, the calculations of the spin stiffness can be improved by directly contracting the TN containing the differentiated terms independent of $v$. The first derivative of the partition function with respect to $v$ introduces an impurity tensor $R$ in the original TN as displayed in Fig.~\ref{fig:stiffness}(a), where
\begin{equation}
R_{n_{1}m_{1},n_{2}m_{2}}^{n_{3}m_{3},n_{4}m_{4}} =\left.\frac{\partial O_v}{\partial v}\right\vert_{v=0}=i(n_1+3m_1)O_{n_{1}m_{1},n_{2}m_{2}}^{n_{3}m_{3},n_{4}m_{4}}.
\end{equation}

Because the partition function is expressed as a tensor product of $O_v$ tensors, the second derivative is composed by two parts: two tensors separately differentiated once and one tensor differentiated twice. The configurations for the second derivative are summed in Fig.~\ref{fig:stiffness}(b), where
\begin{equation}
S_{n_{1}m_{1},n_{2}m_{2}}^{n_{3}m_{3},n_{4}m_{4}} =\left.\frac{\partial^2 O_v}{\partial v^2}\right\vert_{v=0}=-(n_1+3m_1)^2O_{n_{1}m_{1},n_{2}m_{2}}^{n_{3}m_{3},n_{4}m_{4}}.
\end{equation}

In the previous section, we evaluate the expectation value of two-point observables when two impurity tensors locate along a horizontal or vertical line using the fixed point MPS. However, for the spin stiffness, because two $R$ tensors are located at different rows and columns, we cannot simply contract the whole network by squeezing the 2D TN into a 1D chain. Fortunately, another contraction strategy\cite{Vanderstraeten_2015, Vanderstraeten_2016} has been proposed which is based on the channel environment where the linear fixed points can be bent into corner-shaped transfer matrices as the effective ``corner environment" in approximation to infinite quarter-planes as depicted in Fig.~\ref{fig:stiffness}(c). The fixed point equation for the top left corner is shown in Fig.~\ref{fig:stiffness}(d), where the corner-shaped environment is approximated by a bent MPS comprised of two half infinite boundary MPS connected by an corner matrix $K$. Once the corner-shaped fixed points in different quarter-planes are obtained, arbitrary two-point expectation values can be calculated by contracting the network containing $R_i$ and $R_j$ from four corners.  Finally the network is reduced into a shape of cross as shown in Fig.~\ref{fig:stiffness}(e) with the help of the fixed-point eigenvectors $F_L$ and $F_R$.

All contributions of different configurations of the $R$ tensors should be taken into consideration which is achieved by moving the $R$ tensors independently in horizontal and vertical directions. The sum of the infinite geometric series of two observables is a bit more involved. For exponentially decaying observables, long-range observables can be evaluated by inverting the corresponding channel operator as a sum of channel operators
\begin{equation}
\sum_{n=0}^{\infty} \mathbb{T}_L^n \simeq \left[\mathcal{I}-(\mathbb{T}_L-|F_r\rangle\langle F_L|)\right]^{-1},
\end{equation}
where $F_r$ is the right dominant eigenvector of $\mathbb{T}_L$ whose dominant eigenvalue is renormalized to $1$.

The discontinuous jump of the spin stiffness is a characteristic feature for unbinding of topological defects above the BKT transition. In the main text, we show the two-step jump of the superfluid stiffness for both hexatic ($\Delta<1$) and nematic ($\Delta>1$) regime. In contrast to the case for small $\Delta$ where the stiffness develops a cusp point with a small jump at the lower transition temperature due to the unbinding of vortices with charge $q_{\phi} = 1$, in the large $\Delta$ regime the stiffness drops dramatically at the lower transition temperature $T_{c1}$. The different behavior of the stiffness jump at lower transition temperatures between small and large $\Delta$ cases should be related to the dissociations of different kinds of vortices in the $\phi$ and $\theta$ fields. Since the phase twist $v$ is applied to the hexatic $\theta$ field $3$ times larger than to the nematic $\phi$ field, the ratio of the contribution to the total stiffness is roughly $\Delta:9(2-\Delta)$ between $\phi$ and $\theta$ fields for a weak coupling $\lambda$\cite{Touchette_2022}. That is why the first drop of the stiffness is almost $6$ times larger than the second drop upon increasing temperature along $\Delta=1.2$. Such a great drop is different from the strongly coupled case with large $\lambda$ where the topological defects do not result in a jump of the total stiffness but a gradual decrease at the lower crossover. Further investigations into the correlation properties tells us that the extremely weak coherence between vortices in $\theta$ field above $T_{c1}$ should account for the dramatic jump in total stiffness even though the $\theta$ field still remains quasi-long-range ordered (quasi-LRO).

\section{Appendix E: Correlation functions and their correlation lengths}

To reveal the nature of the topological excitations in different phases, we calculate various two-point correlation functions and the corresponding correlation lengths. Here, we present more details for the correlation properties for intra- and inter-component observables. Our results are discussed in the hexatic ($\Delta<1$) and nematic ($\Delta>1$) regimes, separately.

\begin{figure}[thbp]
\centering
\includegraphics[width=.65\linewidth]{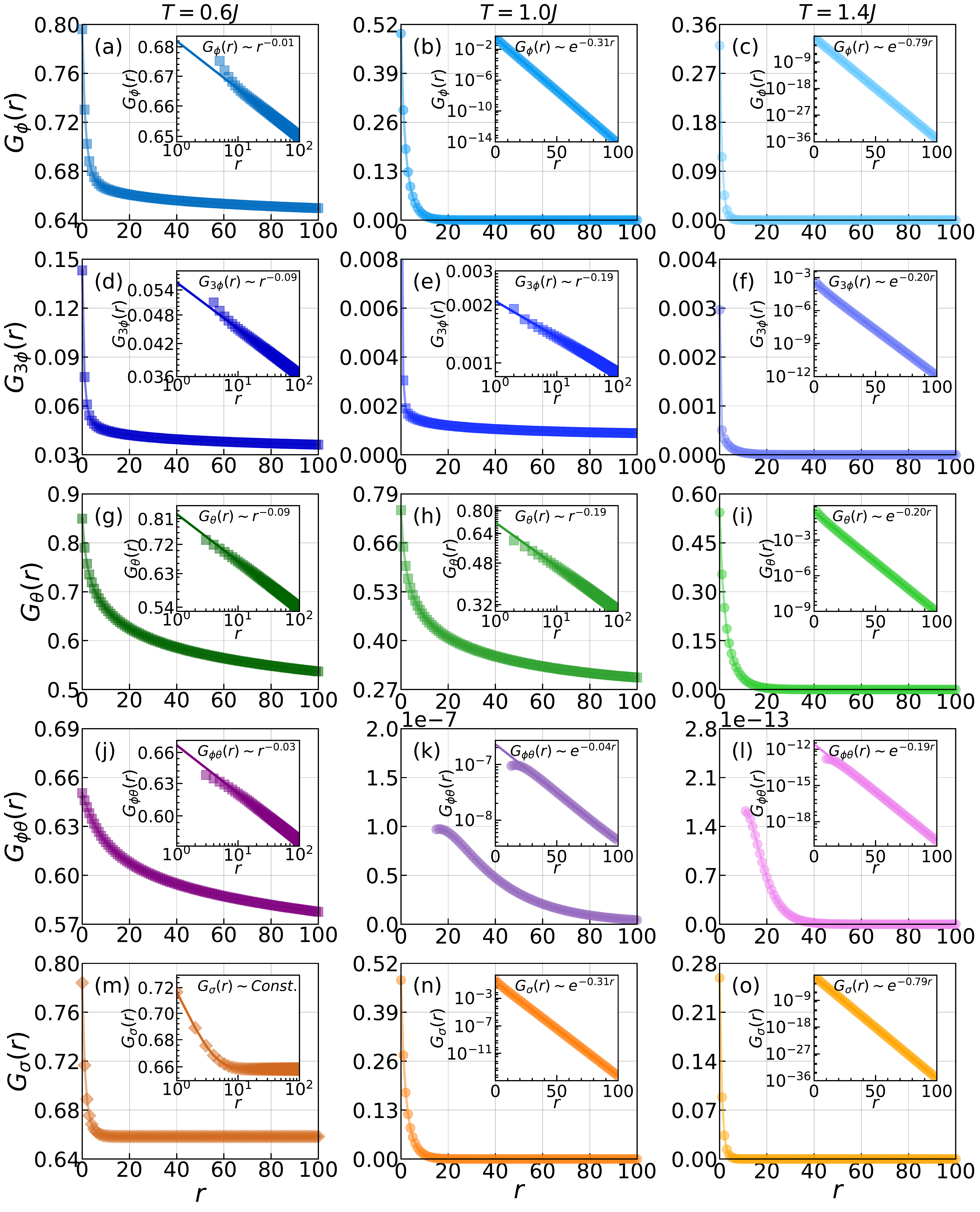}
\caption{
Various two-point correlation functions are calculated under the intra-component coupling ratio $\Delta=0.8$ and the inter-component coupling $\lambda=0.1$. Different columns denote different values of $T=0.6J$, $1.0J$ and $1.4J$ as indicated.}
\label{fig:GsmDelta}
\end{figure}

\begin{figure}[thbp]
\centering
\includegraphics[width=.6\linewidth]{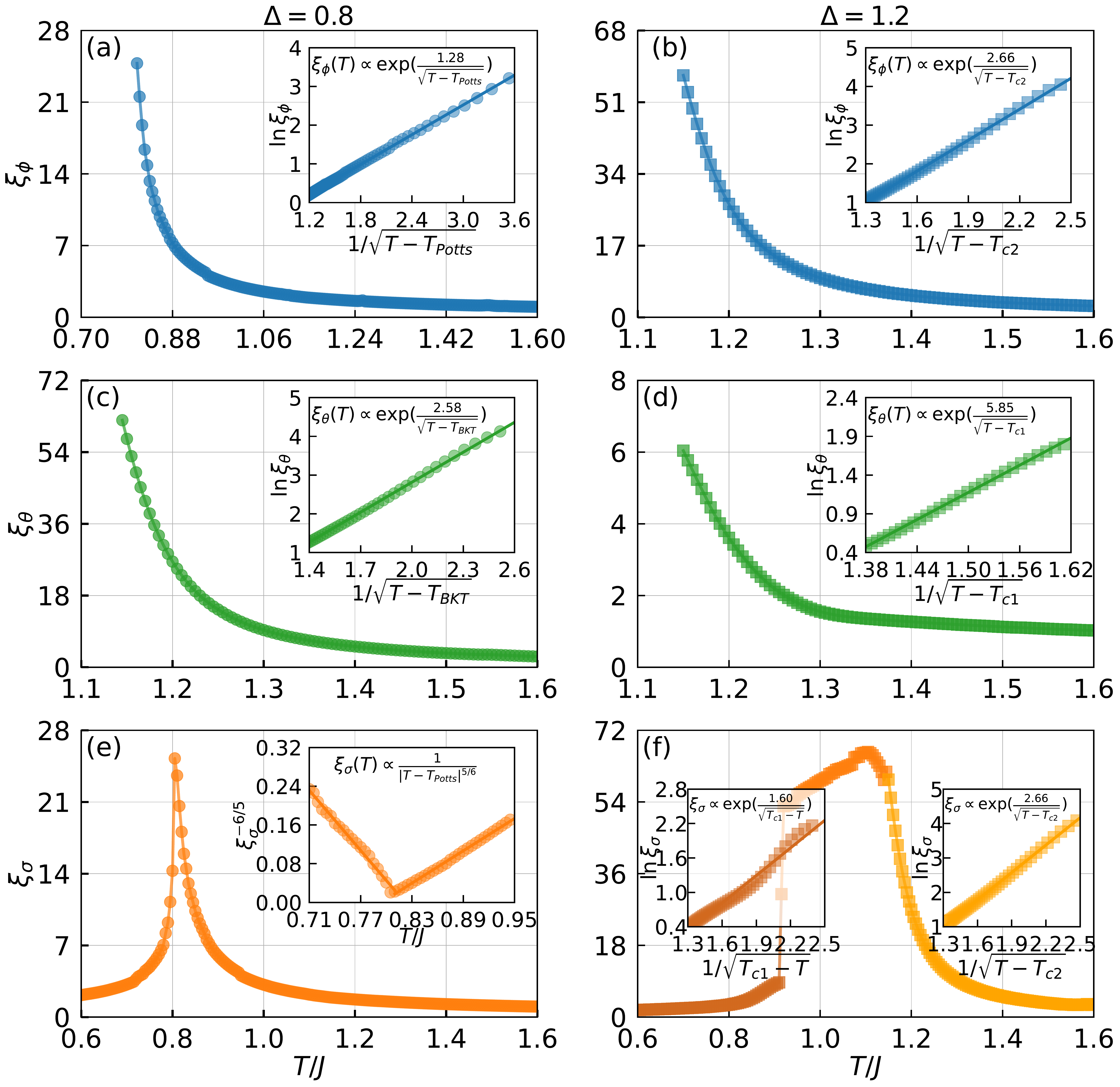}
\caption{
Different correlation lengths $\xi_{\phi}$, $\xi_{\theta}$ and $\xi_{\sigma}$ are calculated for nematic $\phi$, hexatic $\theta$ and intra-component $\sigma$ under $\lambda=0.1$, respectively. Different columns denote different values of $\Delta=0.8$ and $\Delta=1.2$ as indicated.}
\label{fig:clen}
\end{figure}

The correlation functions for $\Delta=0.8$ are summarized in Fig.~\ref{fig:GsmDelta} and the corresponding correlation lengths are displayed in the first column of Fig.~\ref{fig:clen}.

In the $T=0.6J$ column below $T_{c1}$,  the intra-component correlation functions $G_{\phi}(r)$ in Fig.~\ref{fig:GsmDelta} (a) and $G_{\theta} (r)$ in Fig.~\ref{fig:GsmDelta} (g) display the algebraic behavior, indicating the bindings of $q_{\phi}=1$ and $q_{\theta}=1$ vortices in the $\phi$ and $\theta$ fields, respectively. Moreover, the power law behavior of $G_{\phi\theta} (r)$ in Fig.~\ref{fig:GsmDelta} (j) implies that the vortices and anti-vortices also bind in the inter-field. Therefore a fully phase-coherent state of the coupled system is established in the low temperature phase. Besides, the inter-component Potts variable develops a true LRO and the correlation function $G_{\sigma}(r)$ in Fig.~\ref{fig:GsmDelta} (m) becomes a constant for large $r$.

In the $T=1.0J$ column of the intermediate temperatures, the most interesting physics happens in the $\phi$ field. When cooling down the system from high temperature disordered phase, the buildup of correlations in the $\theta$ field first occurs at $T_{c2}$ with an algebraically decaying correlation function $G_{\theta}(r)$ in Fig.~\ref{fig:GsmDelta} (h), corresponding to the binding of hexatic vortices with charge $q_{\theta}=1$.  For the fractional vortex paired phase between $T_{c1}$ and $T_{c2}$, the correlation function $G_{\phi}(r)$ decays exponentially, while the correlation function $G_{3\phi}(r)$ of the fractionalized vortices exhibit an algebraic behavior.

As displayed in Fig.~\ref{fig:GsmDelta}(b) and (e), the direct comparison between difference correlation behavior demonstrates that the integer nematic vortices $q_{\phi}=1$ in the $\phi$ field are fractionalized into fractional nematic vortices with $q_{\phi}=1/3$ due to the presence of the inter-component $Z_3$ degrees of freedom. Since the fractional nematic vortices are pointlike topological defects around which the phase angles of spins wind by $2\pi/3$, each fractional nematic vortices should be at the end of a domain-wall string across which the $\phi$ field twists by $2\pi/3$. The existence of multiple Potts domains destroys the phase coherence in $\phi$ field and leads to disorder in inter-component $\sigma$ variables. As we can see, both $G_{\phi\theta}(r)$ in Fig.~\ref{fig:GsmDelta}(k) and $G_{\sigma}(r)$ in Fig.~\ref{fig:GsmDelta} (n) display the exponential decay. We should point out that the correlation length $\xi _{\phi }$ extracted from $G_{\phi}(r)$ could be fitted by an exponentially divergent form
$\xi (T)\propto \exp (b/\sqrt{T-T_{C}})$
above $T_{c1}$ as shown in Fig.~\ref{fig:clen} (a) with $b>0$ as a key feature of the BKT transition, which implies that the transition at $T_{c1}$  is a hybrid BKT and Potts transition. Furthermore, as shown in Fig.~\ref{fig:clen} (e), the correlation length $\xi_\sigma$ for the inter-component Potts variable extracted from
\begin{equation}
\mathrm{e}^{-r/\xi_{\sigma}}\propto \langle\cos(\sigma_i-\sigma_{i+r})\rangle-\langle\cos(\sigma_i)\rangle\langle\cos(\sigma_{i+r})\rangle
\end{equation}
is well fitted by $\xi_{\sigma}\propto 1/|T-T_{c1}|^{5/6}$ in agreement with 2D Potts universality class.

In the $T=1.4J$ column above $T_{c2}$, all the correlation functions decay exponentially and the system is disordered. When approaching $T_{c2}$ from high temperatures, the relevant correlation length $\xi_\theta$ is extracted and displayed in Fig.~\ref{fig:clen}(c),  which can be well fitted by an exponentially divergent form of the BKT transition.

It is interesting to see that the correlation functions of $G_{\theta}(r)$ and $G_{3\phi}(r)$ share the same exponents while $G_{\phi}(r)$ and $G_{\sigma}(r)$ share the same exponents. Such behavior can be interpreted as another proof of the relevance of the inter-component couplings. Due to the relative locking, the $\theta$ and $3\phi$ fields have the same long-wavelength behavior and the correspondence $q_\theta=3q_\phi$ between vortex charges is established.

\begin{figure}[thbp]
\centering
\includegraphics[width=.65\linewidth]{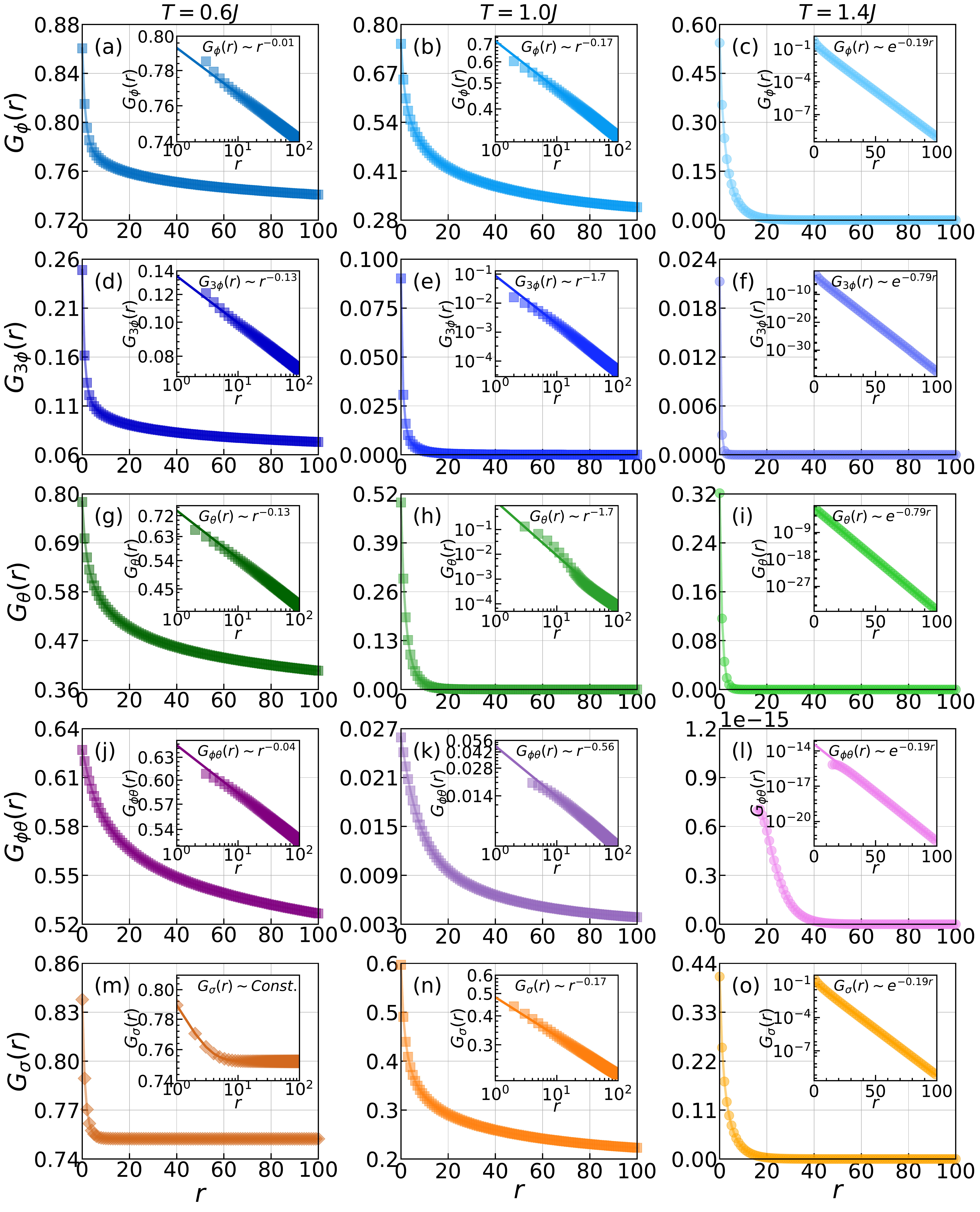}
\caption{
Various two-point correlation functions are calculated under the intra-component coupling ratio $\Delta=1.2$ and the inter-component coupling $\lambda=0.1$. Different columns denote different values of $T=0.6J$, $1.0J$ and $1.4J$ as indicated.}
\label{fig:GlgDelta}
\end{figure}

However, the correlation functions for $\Delta=1.2$ are summarized in Fig.~\ref{fig:GlgDelta} and the corresponding correlation lengths are displayed in the second column of Fig.~\ref{fig:clen}.

In the $T=0.6J$ column below $T_{c1}$, all intra-component correlation functions show a power law behavior and the correlation function for inter-component Potts variable becomes a constant for long distance as displayed in Fig.~\ref{fig:GlgDelta}(m). In fact, the low temperature phases in the hexatic and nematic regime share the same physics.

In the $T=1.0J$ column of the intermediate temperatures, the power law behavior of $G_{\theta}(r)$ in Fig.~\ref{fig:GlgDelta}(h) demonstrates that the quasi-LRO in the $\theta$ field survives through the transition at $T_{c1}$. A direct comparison between the amplitude of $G_{\theta}(r)$ at two sides of $T_{c1}$ in the main text shows that the coherence between spins in the $\theta$ field is greatly suppressed by $3$ orders of magnitude across the transition at $T_{c1}$.  Although the $\theta$ field is still quasi-LRO, the extremely weak coherence above $T_{c1}$ could account for the dramatic jump in total stiffness at $T_{c1}$. The reason for the absence of unbinding transition in the $\theta$ field is that the onset of algebraic correlations in the $\phi$ field in Fig.~\ref{fig:GlgDelta}(b) leads to the quasi-LRO in $\theta$ field because they are coupled by the relevant inter-component term $\lambda\cos(\theta_i-3\phi_i)$. Since the vortices with charge $q_{\phi}=1$ in the $\phi$ field is bound below $T_{c2}$, the dominant topological excitations in the $\theta$ field between $T_{c1}$ and $T_{c2}$ should be composite vortex pairs with $q_{\theta}=3$. Moreover, the composite $q_{\theta}=3$ vortices should further fractionalize into three $q_{\theta}=1$ vortices bound together as a larger extended vortex core to lower the gradient energy cost, especially for weakly coupled cases of smaller $\lambda$. In this way, the correlation $G_{\theta}(r)$ should be viewed as a higher order expansion of the correlation function among composite vortices which has the same long-wavelength behavior as $G_{\phi}(r)$.

In the $T=1.4J$ column above $T_{c2}$, all the correlation functions decay exponentially and the system is disordered. When approaching $T_{c2}$ from high temperatures, the corresponding correlation length $\xi_\phi$ for $G_{\phi}(r)$ as well as $\xi_\theta$ for $G_{\theta}(r)$ are extracted and displayed in Fig.~\ref{fig:clen}(b) and (d),  respectively. Both $\xi_\phi$ and $\xi_\theta$ above $T_{c2}$ can be well fitted by an exponentially divergent form of the BKT transition.

The most interesting physics is revealed in further investigations into the inter-component Potts variable $\sigma_i$. In the strongly coupled case of large $\lambda$,  the LRO of $\sigma$ is found to be lost at $T_{c2}$ together with the destruction of the quasi-LRO in both the hexatic and nematic fields through a BKT transition. However, for the weekly coupled case, the order-disorder transition in $\sigma$ variables splits into two transitions separated by an intermediate phase with Potts quasi-LRO which is regarded as the Potts liquid phase. As shown in Fig.~\ref{fig:GlgDelta}(m), (n) and (o), the correlation function $G_{\sigma}(r)$ exhibits a constant value, an algebraic decay and an exponential decay at $T=0.6J$, $T=1.0J$ and $T=1.4J$, respectively. Such unconventional two-stage phase transitions should be associated with the separated proliferations of the $Z_3$ domain walls and $Z_3$ vortices, similar to two continuous phase transitions in the $Z_p$ clock models with $p > 4$. The correlation length $\xi_{\sigma}$ is also extracted from
$
\mathrm{e}^{-r/\xi_{\sigma}}\propto \langle\cos(\sigma_i-\sigma_{i+r})\rangle-\langle\cos(\sigma_i)\rangle\langle\cos(\sigma_{i+r})\rangle
$
as displayed in Fig.~\ref{fig:clen}(f). The similar behavior of the correlation length was also observed in $p$-state clock models ($p>4$)\cite{Li_2020}. For the clock model, the correlation lengths scale exponentially with $T$ when the quasi-LRO phase is approached from both the ordered and disordered sides, which means that both the transitions at lower and higher temperatures belong to the BKT class. However, as displayed in the inset of Fig.~\ref{fig:clen}(f), the correlation length for $T>T_{c2}$ (right inset) agrees well with the BKT behavior while for $T<T_{c1}$ (left inset) the correlation length cannot fit well in the exponential form. Therefore, it is reasonable to infer that the phase transition at $T_{c1}$ slightly deviates from the standard BKT class.

\section{Appendix F: Topological defects in $Z_3$ variables}

\begin{figure}[thbp]
\centering
\includegraphics[width=.72\linewidth]{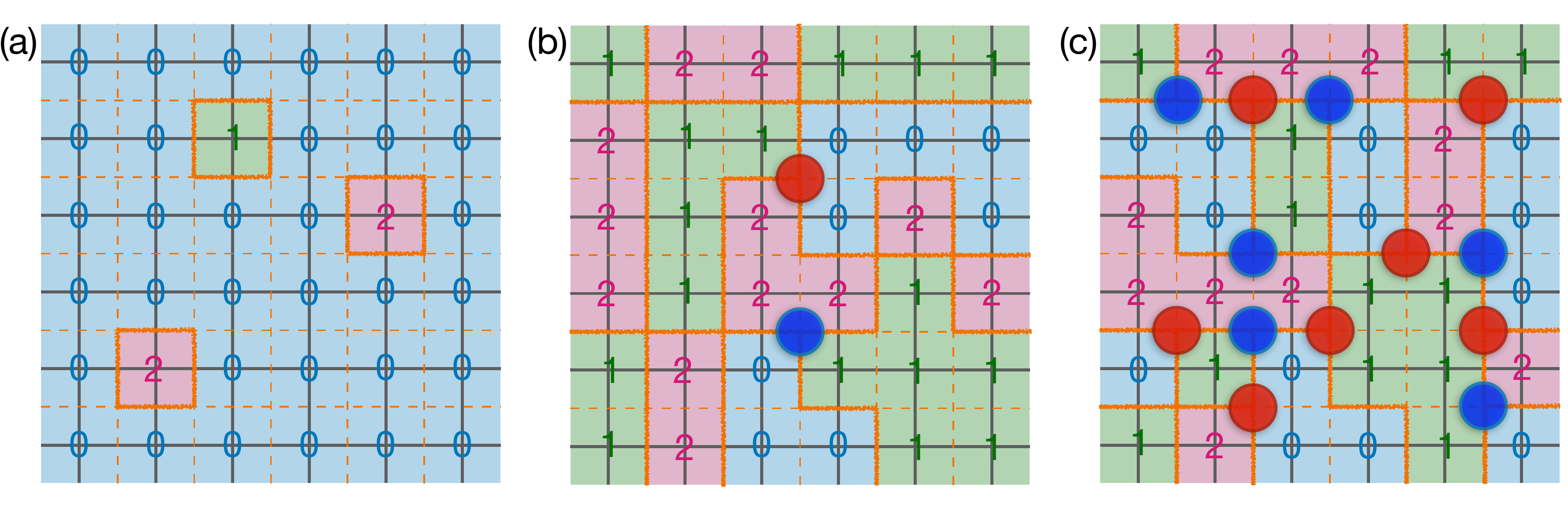}
\caption{
Schematic representations of the topological defects of the $Z_3$ variables in (a) the Potts ordered phase,  (b) the Potts liquid phase, and (c) the Potts disordered phase, respectively. The Potts variables ($\sigma_i=0,1,2$) are define on the original lattice while domain walls (orange thick lines), vortices (red dots), and antivortices (blue dots) are defined on the dual lattice.}
\label{fig:defect}
\end{figure}

To illustrate the splitting of the order-disorder transition in the $Z_3$ Potts variables, the corresponding topological defects are schematic depicted in Fig.~\ref{fig:defect}. Unlike the 2D Ising model where a phase transition is driven by loop-like domain walls, the $Z_3$ Potts model allows the loop-like domain walls as well as $Z_3$ vortices. The $Z_3$ Potts variables $\sigma_i=0,1,2$ are defined at each vertex on a square lattice (black lines). The topological defects, $Z_3$ domain walls and vortices, are defined on the dual lattice (orange dotted line). Each segment of domain wall corresponds to a mismatch in the Potts variables across the edge. The $Z_3$ vortices are determined by the vorticity around the vertex of the dual lattice
\begin{equation}
\omega_p=\frac{1}{3}\sum_{\langle i,j\rangle\in\Box_p}\Delta_{ij}
\end{equation}
where $\Delta_{ij}=\sigma_i-\sigma_j$ are wrapped within $[-1, +1]$ surrounding the plaquette anticlockwise.

For the conventional Potts model, only a single Potts transition is observed because the proliferation of two kinds of topological defects are coupled together. However, we find that a quasi-LRO intermediate phase can appear in the $Z_3$ Potts variables due to the separated excitations of different topological defects in the weakly coupled model. The low temperature transition is driven by the proliferation of domain walls while the high temperature transition is resulted from excitations of free $Z_3$ vortices. The effective core energy of $Z_3$ vortices in the $\sigma$ field of the coupled hexatic-nematic model could be increased by a finite inter-component coupling. Actually, a finite $\lambda$ makes it more costly for the excitations of free vortices in the $\theta$ field which would proliferate through a BKT transition at $T_{c1}$ if $\lambda=0$. On the other hand, for large $\lambda$ limit, the energy of domain walls increases dramatically which forbids the excitations of domain walls before the proliferation of free $Z_3$ vortices and only one transition is left. Since the $\sigma$ variable is determined by $\theta$ and $\phi$ field simultaneously, the relation between the inter-component coupling $\lambda$ and core energy of $Z_3$ vortices is more complicated. An estimation made from an effective long-wavelength model\cite{Touchette_2022} gives that the energies of domain walls are proportional to $\sqrt{\lambda\Delta}$ while the core energies of the vortices in the phase field are proportional to $\Delta$. Therefore a small $\lambda$ is required in the coupled hexatic-nematic model to achieve a relatively high ratio of the core energy of the $Z_3$ vortices compared to domain walls. An investigation of the changes of the phase structure on different $\lambda$ is presented in the main text.

\bibliography{reference}

\begin{thebibliography}{55}
\expandafter\ifx\csname natexlab\endcsname\relax\def\natexlab#1{#1}\fi
\expandafter\ifx\csname bibnamefont\endcsname\relax
  \def\bibnamefont#1{#1}\fi
\expandafter\ifx\csname bibfnamefont\endcsname\relax
  \def\bibfnamefont#1{#1}\fi
\expandafter\ifx\csname citenamefont\endcsname\relax
  \def\citenamefont#1{#1}\fi
\expandafter\ifx\csname url\endcsname\relax
  \def\url#1{\texttt{#1}}\fi
\expandafter\ifx\csname urlprefix\endcsname\relax\def\urlprefix{URL }\fi
\providecommand{\bibinfo}[2]{#2}
\providecommand{\eprint}[2][]{\url{#2}}

\bibitem[{\citenamefont{Berezinsky}(1971)}]{Berezinsky_1971}
\bibinfo{author}{\bibfnamefont{V.}~\bibnamefont{Berezinsky}},
  \bibinfo{journal}{Sov. Phys. JETP} \textbf{\bibinfo{volume}{32}},
  \bibinfo{pages}{493} (\bibinfo{year}{1971}).

\bibitem[{\citenamefont{Kosterlitz and Thouless}(1973)}]{Kosterlitz_1973}
\bibinfo{author}{\bibfnamefont{J.~M.} \bibnamefont{Kosterlitz}}
  \bibnamefont{and} \bibinfo{author}{\bibfnamefont{D.~J.}
  \bibnamefont{Thouless}}, \bibinfo{journal}{Journal of Physics C: Solid State
  Physics} \textbf{\bibinfo{volume}{6}}, \bibinfo{pages}{1181}
  (\bibinfo{year}{1973}), \urlprefix\url{https://doi.org/10.1088}.

\bibitem[{\citenamefont{Kosterlitz}(1974)}]{Kosterlitz_1974}
\bibinfo{author}{\bibfnamefont{J.~M.} \bibnamefont{Kosterlitz}},
  \bibinfo{journal}{Journal of Physics C: Solid State Physics}
  \textbf{\bibinfo{volume}{7}}, \bibinfo{pages}{1046} (\bibinfo{year}{1974}),
  \urlprefix\url{https://doi.org/10.1088}.

\bibitem[{\citenamefont{Halperin and Nelson}(1978)}]{Halperin_1978}
\bibinfo{author}{\bibfnamefont{B.~I.} \bibnamefont{Halperin}} \bibnamefont{and}
  \bibinfo{author}{\bibfnamefont{D.~R.} \bibnamefont{Nelson}},
  \bibinfo{journal}{Phys. Rev. Lett.} \textbf{\bibinfo{volume}{41}},
  \bibinfo{pages}{121} (\bibinfo{year}{1978}),
  \urlprefix\url{https://link.aps.org/doi/10.1103/PhysRevLett.41.121}.

\bibitem[{\citenamefont{Nelson and Halperin}(1979)}]{Nelson_1979}
\bibinfo{author}{\bibfnamefont{D.~R.} \bibnamefont{Nelson}} \bibnamefont{and}
  \bibinfo{author}{\bibfnamefont{B.~I.} \bibnamefont{Halperin}},
  \bibinfo{journal}{Phys. Rev. B} \textbf{\bibinfo{volume}{19}},
  \bibinfo{pages}{2457} (\bibinfo{year}{1979}),
  \urlprefix\url{https://link.aps.org/doi/10.1103/PhysRevB.19.2457}.

\bibitem[{\citenamefont{Young}(1979)}]{Young_1979}
\bibinfo{author}{\bibfnamefont{A.~P.} \bibnamefont{Young}},
  \bibinfo{journal}{Phys. Rev. B} \textbf{\bibinfo{volume}{19}},
  \bibinfo{pages}{1855} (\bibinfo{year}{1979}),
  \urlprefix\url{https://link.aps.org/doi/10.1103/PhysRevB.19.1855}.

\bibitem[{\citenamefont{Kosterlitz}(2016)}]{Kosterlitz_2016}
\bibinfo{author}{\bibfnamefont{J.~M.} \bibnamefont{Kosterlitz}},
  \bibinfo{journal}{Reports on Progress in Physics}
  \textbf{\bibinfo{volume}{79}}, \bibinfo{pages}{026001}
  (\bibinfo{year}{2016}),
  \urlprefix\url{https://dx.doi.org/10.1088/0034-4885/79/2/026001}.

\bibitem[{\citenamefont{Ryzhov et~al.}(2017)\citenamefont{Ryzhov, Tareyeva,
  Fomin, and Tsiok}}]{Ryzhov_2017}
\bibinfo{author}{\bibfnamefont{V.~N.} \bibnamefont{Ryzhov}},
  \bibinfo{author}{\bibfnamefont{E.~E.} \bibnamefont{Tareyeva}},
  \bibinfo{author}{\bibfnamefont{Y.~D.} \bibnamefont{Fomin}}, \bibnamefont{and}
  \bibinfo{author}{\bibfnamefont{E.~N.} \bibnamefont{Tsiok}},
  \bibinfo{journal}{Physics-Uspekhi} \textbf{\bibinfo{volume}{60}},
  \bibinfo{pages}{857} (\bibinfo{year}{2017}),
  \urlprefix\url{https://dx.doi.org/10.3367/UFNe.2017.06.038161}.

\bibitem[{\citenamefont{Strandburg}(1988)}]{Strandburg_1988}
\bibinfo{author}{\bibfnamefont{K.~J.} \bibnamefont{Strandburg}},
  \bibinfo{journal}{Rev. Mod. Phys.} \textbf{\bibinfo{volume}{60}},
  \bibinfo{pages}{161} (\bibinfo{year}{1988}),
  \urlprefix\url{https://link.aps.org/doi/10.1103/RevModPhys.60.161}.

\bibitem[{\citenamefont{Bernard and Krauth}(2011)}]{Bernard_2011}
\bibinfo{author}{\bibfnamefont{E.~P.} \bibnamefont{Bernard}} \bibnamefont{and}
  \bibinfo{author}{\bibfnamefont{W.}~\bibnamefont{Krauth}},
  \bibinfo{journal}{Phys. Rev. Lett.} \textbf{\bibinfo{volume}{107}},
  \bibinfo{pages}{155704} (\bibinfo{year}{2011}),
  \urlprefix\url{https://link.aps.org/doi/10.1103/PhysRevLett.107.155704}.

\bibitem[{\citenamefont{Chae et~al.}(2012)\citenamefont{Chae, Lee, Horibe,
  Tanimura, Mori, Gao, Carr, and Cheong}}]{Chae_2012}
\bibinfo{author}{\bibfnamefont{S.~C.} \bibnamefont{Chae}},
  \bibinfo{author}{\bibfnamefont{N.}~\bibnamefont{Lee}},
  \bibinfo{author}{\bibfnamefont{Y.}~\bibnamefont{Horibe}},
  \bibinfo{author}{\bibfnamefont{M.}~\bibnamefont{Tanimura}},
  \bibinfo{author}{\bibfnamefont{S.}~\bibnamefont{Mori}},
  \bibinfo{author}{\bibfnamefont{B.}~\bibnamefont{Gao}},
  \bibinfo{author}{\bibfnamefont{S.}~\bibnamefont{Carr}}, \bibnamefont{and}
  \bibinfo{author}{\bibfnamefont{S.-W.} \bibnamefont{Cheong}},
  \bibinfo{journal}{Phys. Rev. Lett.} \textbf{\bibinfo{volume}{108}},
  \bibinfo{pages}{167603} (\bibinfo{year}{2012}),
  \urlprefix\url{https://link.aps.org/doi/10.1103/PhysRevLett.108.167603}.

\bibitem[{\citenamefont{Radzihovsky et~al.}(2008)\citenamefont{Radzihovsky,
  Weichman, and Park}}]{Radzihovsky_2008}
\bibinfo{author}{\bibfnamefont{L.}~\bibnamefont{Radzihovsky}},
  \bibinfo{author}{\bibfnamefont{P.~B.} \bibnamefont{Weichman}},
  \bibnamefont{and} \bibinfo{author}{\bibfnamefont{J.~I.} \bibnamefont{Park}},
  \bibinfo{journal}{Annals of Physics} \textbf{\bibinfo{volume}{323}},
  \bibinfo{pages}{2376} (\bibinfo{year}{2008}), ISSN \bibinfo{issn}{0003-4916},
  \urlprefix\url{https://www.sciencedirect.com/science/article/pii/S0003491608000742}.

\bibitem[{\citenamefont{Shahbazi and Ghanbari}(2006)}]{Shahbazi_2006}
\bibinfo{author}{\bibfnamefont{F.}~\bibnamefont{Shahbazi}} \bibnamefont{and}
  \bibinfo{author}{\bibfnamefont{R.}~\bibnamefont{Ghanbari}},
  \bibinfo{journal}{Phys. Rev. E} \textbf{\bibinfo{volume}{74}},
  \bibinfo{pages}{021705} (\bibinfo{year}{2006}),
  \urlprefix\url{https://link.aps.org/doi/10.1103/PhysRevE.74.021705}.

\bibitem[{\citenamefont{de~Parny et~al.}(2016)\citenamefont{de~Parny,
  Ran\ifmmode~\mbox{\c{c}}\else \c{c}\fi{}on, and Roscilde}}]{Parny_2016}
\bibinfo{author}{\bibfnamefont{L.~d.~F.} \bibnamefont{de~Parny}},
  \bibinfo{author}{\bibfnamefont{A.}~\bibnamefont{Ran\ifmmode~\mbox{\c{c}}\else
  \c{c}\fi{}on}}, \bibnamefont{and}
  \bibinfo{author}{\bibfnamefont{T.}~\bibnamefont{Roscilde}},
  \bibinfo{journal}{Phys. Rev. A} \textbf{\bibinfo{volume}{93}},
  \bibinfo{pages}{023639} (\bibinfo{year}{2016}),
  \urlprefix\url{https://link.aps.org/doi/10.1103/PhysRevA.93.023639}.

\bibitem[{\citenamefont{Serna et~al.}(2017)\citenamefont{Serna, Chalker, and
  Fendley}}]{Serna_2017}
\bibinfo{author}{\bibfnamefont{P.}~\bibnamefont{Serna}},
  \bibinfo{author}{\bibfnamefont{J.~T.} \bibnamefont{Chalker}},
  \bibnamefont{and} \bibinfo{author}{\bibfnamefont{P.}~\bibnamefont{Fendley}},
  \bibinfo{journal}{Journal of Physics A: Mathematical and Theoretical}
  \textbf{\bibinfo{volume}{50}}, \bibinfo{pages}{424003}
  (\bibinfo{year}{2017}),
  \urlprefix\url{https://dx.doi.org/10.1088/1751-8121/aa89a1}.

\bibitem[{\citenamefont{Kobayashi et~al.}(2019)\citenamefont{Kobayashi, Eto,
  and Nitta}}]{Kobayashi_2019}
\bibinfo{author}{\bibfnamefont{M.}~\bibnamefont{Kobayashi}},
  \bibinfo{author}{\bibfnamefont{M.}~\bibnamefont{Eto}}, \bibnamefont{and}
  \bibinfo{author}{\bibfnamefont{M.}~\bibnamefont{Nitta}},
  \bibinfo{journal}{Phys. Rev. Lett.} \textbf{\bibinfo{volume}{123}},
  \bibinfo{pages}{075303} (\bibinfo{year}{2019}),
  \urlprefix\url{https://link.aps.org/doi/10.1103/PhysRevLett.123.075303}.

\bibitem[{\citenamefont{Bruinsma and Aeppli}(1982)}]{Bruinsma_1982}
\bibinfo{author}{\bibfnamefont{R.}~\bibnamefont{Bruinsma}} \bibnamefont{and}
  \bibinfo{author}{\bibfnamefont{G.}~\bibnamefont{Aeppli}},
  \bibinfo{journal}{Phys. Rev. Lett.} \textbf{\bibinfo{volume}{48}},
  \bibinfo{pages}{1625} (\bibinfo{year}{1982}),
  \urlprefix\url{https://link.aps.org/doi/10.1103/PhysRevLett.48.1625}.

\bibitem[{\citenamefont{Aeppli and Bruinsma}(1984)}]{Aeppli_1984}
\bibinfo{author}{\bibfnamefont{G.}~\bibnamefont{Aeppli}} \bibnamefont{and}
  \bibinfo{author}{\bibfnamefont{R.}~\bibnamefont{Bruinsma}},
  \bibinfo{journal}{Phys. Rev. Lett.} \textbf{\bibinfo{volume}{53}},
  \bibinfo{pages}{2133} (\bibinfo{year}{1984}),
  \urlprefix\url{https://link.aps.org/doi/10.1103/PhysRevLett.53.2133}.

\bibitem[{\citenamefont{Jiang et~al.}(1993)\citenamefont{Jiang, Huang, Ko,
  Stoebe, Jin, and Huang}}]{Jiang_1993}
\bibinfo{author}{\bibfnamefont{I.~M.} \bibnamefont{Jiang}},
  \bibinfo{author}{\bibfnamefont{S.~N.} \bibnamefont{Huang}},
  \bibinfo{author}{\bibfnamefont{J.~Y.} \bibnamefont{Ko}},
  \bibinfo{author}{\bibfnamefont{T.}~\bibnamefont{Stoebe}},
  \bibinfo{author}{\bibfnamefont{A.~J.} \bibnamefont{Jin}}, \bibnamefont{and}
  \bibinfo{author}{\bibfnamefont{C.~C.} \bibnamefont{Huang}},
  \bibinfo{journal}{Phys. Rev. E} \textbf{\bibinfo{volume}{48}},
  \bibinfo{pages}{R3240} (\bibinfo{year}{1993}),
  \urlprefix\url{https://link.aps.org/doi/10.1103/PhysRevE.48.R3240}.

\bibitem[{\citenamefont{Jiang et~al.}(1996)\citenamefont{Jiang, Stoebe, and
  Huang}}]{Jiang_1996}
\bibinfo{author}{\bibfnamefont{I.~M.} \bibnamefont{Jiang}},
  \bibinfo{author}{\bibfnamefont{T.}~\bibnamefont{Stoebe}}, \bibnamefont{and}
  \bibinfo{author}{\bibfnamefont{C.~C.} \bibnamefont{Huang}},
  \bibinfo{journal}{Phys. Rev. Lett.} \textbf{\bibinfo{volume}{76}},
  \bibinfo{pages}{2910} (\bibinfo{year}{1996}),
  \urlprefix\url{https://link.aps.org/doi/10.1103/PhysRevLett.76.2910}.

\bibitem[{\citenamefont{Drouin-Touchette
  et~al.}(2022)\citenamefont{Drouin-Touchette, Orth, Coleman, Chandra, and
  Lubensky}}]{Touchette_2022}
\bibinfo{author}{\bibfnamefont{V.}~\bibnamefont{Drouin-Touchette}},
  \bibinfo{author}{\bibfnamefont{P.~P.} \bibnamefont{Orth}},
  \bibinfo{author}{\bibfnamefont{P.}~\bibnamefont{Coleman}},
  \bibinfo{author}{\bibfnamefont{P.}~\bibnamefont{Chandra}}, \bibnamefont{and}
  \bibinfo{author}{\bibfnamefont{T.~C.} \bibnamefont{Lubensky}},
  \bibinfo{journal}{Phys. Rev. X} \textbf{\bibinfo{volume}{12}},
  \bibinfo{pages}{011043} (\bibinfo{year}{2022}),
  \urlprefix\url{https://link.aps.org/doi/10.1103/PhysRevX.12.011043}.

\bibitem[{\citenamefont{Verstraete et~al.}(2008)\citenamefont{Verstraete, Murg,
  and Cirac}}]{Verstraete_2008}
\bibinfo{author}{\bibfnamefont{F.}~\bibnamefont{Verstraete}},
  \bibinfo{author}{\bibfnamefont{V.}~\bibnamefont{Murg}}, \bibnamefont{and}
  \bibinfo{author}{\bibfnamefont{J.}~\bibnamefont{Cirac}},
  \bibinfo{journal}{Advances in Physics} \textbf{\bibinfo{volume}{57}},
  \bibinfo{pages}{143} (\bibinfo{year}{2008}),
  \eprint{https://doi.org/10.1080/14789940801912366},
  \urlprefix\url{https://doi.org/10.1080/14789940801912366}.

\bibitem[{\citenamefont{Or{\'u}s}(2014)}]{Orus_2014}
\bibinfo{author}{\bibfnamefont{R.}~\bibnamefont{Or{\'u}s}},
  \bibinfo{journal}{Annals of Physics} \textbf{\bibinfo{volume}{349}},
  \bibinfo{pages}{117 } (\bibinfo{year}{2014}), ISSN \bibinfo{issn}{0003-4916},
  \urlprefix\url{http://www.sciencedirect.com/science/article/pii/S0003491614001596}.

\bibitem[{\citenamefont{Haegeman and
  Verstraete}(2017)}]{Haegeman-Verstraete2017}
\bibinfo{author}{\bibfnamefont{J.}~\bibnamefont{Haegeman}} \bibnamefont{and}
  \bibinfo{author}{\bibfnamefont{F.}~\bibnamefont{Verstraete}},
  \bibinfo{journal}{Annual Review of Condensed Matter Physics}
  \textbf{\bibinfo{volume}{8}}, \bibinfo{pages}{355} (\bibinfo{year}{2017}),
  \eprint{https://doi.org/10.1146/annurev-conmatphys-031016-025507},
  \urlprefix\url{https://doi.org/10.1146/annurev-conmatphys-031016-025507}.

\bibitem[{\citenamefont{Zauner-Stauber
  et~al.}(2018)\citenamefont{Zauner-Stauber, Vanderstraeten, Fishman,
  Verstraete, and Haegeman}}]{VUMPS}
\bibinfo{author}{\bibfnamefont{V.}~\bibnamefont{Zauner-Stauber}},
  \bibinfo{author}{\bibfnamefont{L.}~\bibnamefont{Vanderstraeten}},
  \bibinfo{author}{\bibfnamefont{M.~T.} \bibnamefont{Fishman}},
  \bibinfo{author}{\bibfnamefont{F.}~\bibnamefont{Verstraete}},
  \bibnamefont{and} \bibinfo{author}{\bibfnamefont{J.}~\bibnamefont{Haegeman}},
  \bibinfo{journal}{Phys. Rev. B} \textbf{\bibinfo{volume}{97}},
  \bibinfo{pages}{045145} (\bibinfo{year}{2018}),
  \urlprefix\url{https://link.aps.org/doi/10.1103/PhysRevB.97.045145}.

\bibitem[{\citenamefont{Fishman et~al.}(2018)\citenamefont{Fishman,
  Vanderstraeten, Zauner-Stauber, Haegeman, and Verstraete}}]{Fishman_2018}
\bibinfo{author}{\bibfnamefont{M.~T.} \bibnamefont{Fishman}},
  \bibinfo{author}{\bibfnamefont{L.}~\bibnamefont{Vanderstraeten}},
  \bibinfo{author}{\bibfnamefont{V.}~\bibnamefont{Zauner-Stauber}},
  \bibinfo{author}{\bibfnamefont{J.}~\bibnamefont{Haegeman}}, \bibnamefont{and}
  \bibinfo{author}{\bibfnamefont{F.}~\bibnamefont{Verstraete}},
  \bibinfo{journal}{Phys. Rev. B} \textbf{\bibinfo{volume}{98}},
  \bibinfo{pages}{235148} (\bibinfo{year}{2018}),
  \urlprefix\url{https://link.aps.org/doi/10.1103/PhysRevB.98.235148}.

\bibitem[{\citenamefont{Vanderstraeten
  et~al.}(2019{\natexlab{a}})\citenamefont{Vanderstraeten, Haegeman, and
  Verstraete}}]{Laurens_Haegeman_2019}
\bibinfo{author}{\bibfnamefont{L.}~\bibnamefont{Vanderstraeten}},
  \bibinfo{author}{\bibfnamefont{J.}~\bibnamefont{Haegeman}}, \bibnamefont{and}
  \bibinfo{author}{\bibfnamefont{F.}~\bibnamefont{Verstraete}},
  \bibinfo{journal}{SciPost Phys. Lect. Notes} p.~\bibinfo{pages}{7}
  (\bibinfo{year}{2019}{\natexlab{a}}),
  \urlprefix\url{https://scipost.org/10.21468/SciPostPhysLectNotes.7}.

\bibitem[{\citenamefont{Vanderstraeten
  et~al.}(2019{\natexlab{b}})\citenamefont{Vanderstraeten, Vanhecke, L\"auchli,
  and Verstraete}}]{Laurens_Bram_2019}
\bibinfo{author}{\bibfnamefont{L.}~\bibnamefont{Vanderstraeten}},
  \bibinfo{author}{\bibfnamefont{B.}~\bibnamefont{Vanhecke}},
  \bibinfo{author}{\bibfnamefont{A.~M.} \bibnamefont{L\"auchli}},
  \bibnamefont{and}
  \bibinfo{author}{\bibfnamefont{F.}~\bibnamefont{Verstraete}},
  \bibinfo{journal}{Phys. Rev. E} \textbf{\bibinfo{volume}{100}},
  \bibinfo{pages}{062136} (\bibinfo{year}{2019}{\natexlab{b}}),
  \urlprefix\url{https://link.aps.org/doi/10.1103/PhysRevE.100.062136}.

\bibitem[{\citenamefont{Li et~al.}(2020)\citenamefont{Li, Yang, Xie, Tu, Liao,
  and Xiang}}]{Li_2020}
\bibinfo{author}{\bibfnamefont{Z.-Q.} \bibnamefont{Li}},
  \bibinfo{author}{\bibfnamefont{L.-P.} \bibnamefont{Yang}},
  \bibinfo{author}{\bibfnamefont{Z.~Y.} \bibnamefont{Xie}},
  \bibinfo{author}{\bibfnamefont{H.-H.} \bibnamefont{Tu}},
  \bibinfo{author}{\bibfnamefont{H.-J.} \bibnamefont{Liao}}, \bibnamefont{and}
  \bibinfo{author}{\bibfnamefont{T.}~\bibnamefont{Xiang}},
  \bibinfo{journal}{Phys. Rev. E} \textbf{\bibinfo{volume}{101}},
  \bibinfo{pages}{060105} (\bibinfo{year}{2020}),
  \urlprefix\url{https://link.aps.org/doi/10.1103/PhysRevE.101.060105}.

\bibitem[{\citenamefont{Song and Zhang}(2022{\natexlab{a}})}]{Song_2022_2}
\bibinfo{author}{\bibfnamefont{F.-F.} \bibnamefont{Song}} \bibnamefont{and}
  \bibinfo{author}{\bibfnamefont{G.-M.} \bibnamefont{Zhang}},
  \bibinfo{journal}{Phys. Rev. Lett.} \textbf{\bibinfo{volume}{128}},
  \bibinfo{pages}{195301} (\bibinfo{year}{2022}{\natexlab{a}}),
  \urlprefix\url{https://link.aps.org/doi/10.1103/PhysRevLett.128.195301}.

\bibitem[{\citenamefont{Granato and Kosterlitz}(1986)}]{Grannato_1986}
\bibinfo{author}{\bibfnamefont{E.}~\bibnamefont{Granato}} \bibnamefont{and}
  \bibinfo{author}{\bibfnamefont{J.~M.} \bibnamefont{Kosterlitz}},
  \bibinfo{journal}{Phys. Rev. B} \textbf{\bibinfo{volume}{33}},
  \bibinfo{pages}{4767} (\bibinfo{year}{1986}),
  \urlprefix\url{https://link.aps.org/doi/10.1103/PhysRevB.33.4767}.

\bibitem[{\citenamefont{Poderoso et~al.}(2011)\citenamefont{Poderoso, Arenzon,
  and Levin}}]{Poderoso_2011}
\bibinfo{author}{\bibfnamefont{F.~C.} \bibnamefont{Poderoso}},
  \bibinfo{author}{\bibfnamefont{J.~J.} \bibnamefont{Arenzon}},
  \bibnamefont{and} \bibinfo{author}{\bibfnamefont{Y.}~\bibnamefont{Levin}},
  \bibinfo{journal}{Phys. Rev. Lett.} \textbf{\bibinfo{volume}{106}},
  \bibinfo{pages}{067202} (\bibinfo{year}{2011}),
  \urlprefix\url{https://link.aps.org/doi/10.1103/PhysRevLett.106.067202}.

\bibitem[{\citenamefont{Canova et~al.}(2014)\citenamefont{Canova, Levin, and
  Arenzon}}]{Canova_2014}
\bibinfo{author}{\bibfnamefont{G.~A.} \bibnamefont{Canova}},
  \bibinfo{author}{\bibfnamefont{Y.}~\bibnamefont{Levin}}, \bibnamefont{and}
  \bibinfo{author}{\bibfnamefont{J.~J.} \bibnamefont{Arenzon}},
  \bibinfo{journal}{Phys. Rev. E} \textbf{\bibinfo{volume}{89}},
  \bibinfo{pages}{012126} (\bibinfo{year}{2014}),
  \urlprefix\url{https://link.aps.org/doi/10.1103/PhysRevE.89.012126}.

\bibitem[{\citenamefont{Canova et~al.}(2016)\citenamefont{Canova, Levin, and
  Arenzon}}]{Canova_2016}
\bibinfo{author}{\bibfnamefont{G.~A.} \bibnamefont{Canova}},
  \bibinfo{author}{\bibfnamefont{Y.}~\bibnamefont{Levin}}, \bibnamefont{and}
  \bibinfo{author}{\bibfnamefont{J.~J.} \bibnamefont{Arenzon}},
  \bibinfo{journal}{Phys. Rev. E} \textbf{\bibinfo{volume}{94}},
  \bibinfo{pages}{032140} (\bibinfo{year}{2016}),
  \urlprefix\url{https://link.aps.org/doi/10.1103/PhysRevE.94.032140}.

\bibitem[{\citenamefont{Song and Zhang}(2021)}]{Song_2021}
\bibinfo{author}{\bibfnamefont{F.-F.} \bibnamefont{Song}} \bibnamefont{and}
  \bibinfo{author}{\bibfnamefont{G.-M.} \bibnamefont{Zhang}},
  \bibinfo{journal}{Phys. Rev. B} \textbf{\bibinfo{volume}{103}},
  \bibinfo{pages}{024518} (\bibinfo{year}{2021}),
  \urlprefix\url{https://link.aps.org/doi/10.1103/PhysRevB.103.024518}.

\bibitem[{\citenamefont{Song and Zhang}(2022{\natexlab{b}})}]{Song_2022}
\bibinfo{author}{\bibfnamefont{F.-F.} \bibnamefont{Song}} \bibnamefont{and}
  \bibinfo{author}{\bibfnamefont{G.-M.} \bibnamefont{Zhang}},
  \bibinfo{journal}{Phys. Rev. B} \textbf{\bibinfo{volume}{105}},
  \bibinfo{pages}{134516} (\bibinfo{year}{2022}{\natexlab{b}}),
  \urlprefix\url{https://link.aps.org/doi/10.1103/PhysRevB.105.134516}.

\bibitem[{\citenamefont{Vidal et~al.}(2003)\citenamefont{Vidal, Latorre, Rico,
  and Kitaev}}]{Vidal_2003}
\bibinfo{author}{\bibfnamefont{G.}~\bibnamefont{Vidal}},
  \bibinfo{author}{\bibfnamefont{J.~I.} \bibnamefont{Latorre}},
  \bibinfo{author}{\bibfnamefont{E.}~\bibnamefont{Rico}}, \bibnamefont{and}
  \bibinfo{author}{\bibfnamefont{A.}~\bibnamefont{Kitaev}},
  \bibinfo{journal}{Phys. Rev. Lett.} \textbf{\bibinfo{volume}{90}},
  \bibinfo{pages}{227902} (\bibinfo{year}{2003}),
  \urlprefix\url{https://link.aps.org/doi/10.1103/PhysRevLett.90.227902}.

\bibitem[{\citenamefont{Pollmann et~al.}(2009)\citenamefont{Pollmann, Mukerjee,
  Turner, and Moore}}]{Pollmann_2009}
\bibinfo{author}{\bibfnamefont{F.}~\bibnamefont{Pollmann}},
  \bibinfo{author}{\bibfnamefont{S.}~\bibnamefont{Mukerjee}},
  \bibinfo{author}{\bibfnamefont{A.~M.} \bibnamefont{Turner}},
  \bibnamefont{and} \bibinfo{author}{\bibfnamefont{J.~E.} \bibnamefont{Moore}},
  \bibinfo{journal}{Phys. Rev. Lett.} \textbf{\bibinfo{volume}{102}},
  \bibinfo{pages}{255701} (\bibinfo{year}{2009}),
  \urlprefix\url{https://link.aps.org/doi/10.1103/PhysRevLett.102.255701}.

\bibitem[{SM(See Supplementary Materials for the tensor network method and
  detailed numerical calculations.)}]{SM}
 (\bibinfo{year}{See Supplementary Materials for the tensor network method and
  detailed numerical calculations.}).

\bibitem[{\citenamefont{H\"ubscher and Wessel}(2013)}]{Hubscher_2013}
\bibinfo{author}{\bibfnamefont{D.~M.} \bibnamefont{H\"ubscher}}
  \bibnamefont{and} \bibinfo{author}{\bibfnamefont{S.}~\bibnamefont{Wessel}},
  \bibinfo{journal}{Phys. Rev. E} \textbf{\bibinfo{volume}{87}},
  \bibinfo{pages}{062112} (\bibinfo{year}{2013}),
  \urlprefix\url{https://link.aps.org/doi/10.1103/PhysRevE.87.062112}.

\bibitem[{\citenamefont{Wu}(1982)}]{Wu_1982}
\bibinfo{author}{\bibfnamefont{F.~Y.} \bibnamefont{Wu}}, \bibinfo{journal}{Rev.
  Mod. Phys.} \textbf{\bibinfo{volume}{54}}, \bibinfo{pages}{235}
  (\bibinfo{year}{1982}),
  \urlprefix\url{https://link.aps.org/doi/10.1103/RevModPhys.54.235}.

\bibitem[{\citenamefont{Einhorn et~al.}(1980)\citenamefont{Einhorn, Savit, and
  Rabinovici}}]{Einhorn-1980}
\bibinfo{author}{\bibfnamefont{M.~B.} \bibnamefont{Einhorn}},
  \bibinfo{author}{\bibfnamefont{R.}~\bibnamefont{Savit}}, \bibnamefont{and}
  \bibinfo{author}{\bibfnamefont{E.}~\bibnamefont{Rabinovici}},
  \bibinfo{journal}{Nuclear Physics B} \textbf{\bibinfo{volume}{170}},
  \bibinfo{pages}{16} (\bibinfo{year}{1980}), ISSN \bibinfo{issn}{0550-3213},
  \urlprefix\url{https://www.sciencedirect.com/science/article/pii/0550321380904733}.

\bibitem[{\citenamefont{Bhattacharya and Ray}(2016)}]{Bhattacharya_2016}
\bibinfo{author}{\bibfnamefont{S.}~\bibnamefont{Bhattacharya}}
  \bibnamefont{and} \bibinfo{author}{\bibfnamefont{P.}~\bibnamefont{Ray}},
  \bibinfo{journal}{Phys. Rev. Lett.} \textbf{\bibinfo{volume}{116}},
  \bibinfo{pages}{097206} (\bibinfo{year}{2016}),
  \urlprefix\url{https://link.aps.org/doi/10.1103/PhysRevLett.116.097206}.

\bibitem[{\citenamefont{Pindak et~al.}(1981)\citenamefont{Pindak, Moncton,
  Davey, and Goodby}}]{Pindak_1981}
\bibinfo{author}{\bibfnamefont{R.}~\bibnamefont{Pindak}},
  \bibinfo{author}{\bibfnamefont{D.~E.} \bibnamefont{Moncton}},
  \bibinfo{author}{\bibfnamefont{S.~C.} \bibnamefont{Davey}}, \bibnamefont{and}
  \bibinfo{author}{\bibfnamefont{J.~W.} \bibnamefont{Goodby}},
  \bibinfo{journal}{Phys. Rev. Lett.} \textbf{\bibinfo{volume}{46}},
  \bibinfo{pages}{1135} (\bibinfo{year}{1981}),
  \urlprefix\url{https://link.aps.org/doi/10.1103/PhysRevLett.46.1135}.

\bibitem[{\citenamefont{Huang et~al.}(1981)\citenamefont{Huang, Viner, Pindak,
  and Goodby}}]{Huang_1981}
\bibinfo{author}{\bibfnamefont{C.~C.} \bibnamefont{Huang}},
  \bibinfo{author}{\bibfnamefont{J.~M.} \bibnamefont{Viner}},
  \bibinfo{author}{\bibfnamefont{R.}~\bibnamefont{Pindak}}, \bibnamefont{and}
  \bibinfo{author}{\bibfnamefont{J.~W.} \bibnamefont{Goodby}},
  \bibinfo{journal}{Phys. Rev. Lett.} \textbf{\bibinfo{volume}{46}},
  \bibinfo{pages}{1289} (\bibinfo{year}{1981}),
  \urlprefix\url{https://link.aps.org/doi/10.1103/PhysRevLett.46.1289}.

\bibitem[{\citenamefont{Chou et~al.}(1998)\citenamefont{Chou, Jin, Hui, Huang,
  and Ho}}]{Chou_1998}
\bibinfo{author}{\bibfnamefont{C.-F.} \bibnamefont{Chou}},
  \bibinfo{author}{\bibfnamefont{A.~J.} \bibnamefont{Jin}},
  \bibinfo{author}{\bibfnamefont{S.~W.} \bibnamefont{Hui}},
  \bibinfo{author}{\bibfnamefont{C.~C.} \bibnamefont{Huang}}, \bibnamefont{and}
  \bibinfo{author}{\bibfnamefont{J.~T.} \bibnamefont{Ho}},
  \bibinfo{journal}{Science} \textbf{\bibinfo{volume}{280}},
  \bibinfo{pages}{1424} (\bibinfo{year}{1998}),
  \urlprefix\url{https://www.science.org/doi/abs/10.1126/science.280.5368.1424}.

\bibitem[{\citenamefont{Nelson and Halperin}(1980)}]{Nelson_1980}
\bibinfo{author}{\bibfnamefont{D.~R.} \bibnamefont{Nelson}} \bibnamefont{and}
  \bibinfo{author}{\bibfnamefont{B.~I.} \bibnamefont{Halperin}},
  \bibinfo{journal}{Phys. Rev. B} \textbf{\bibinfo{volume}{21}},
  \bibinfo{pages}{5312} (\bibinfo{year}{1980}),
  \urlprefix\url{https://link.aps.org/doi/10.1103/PhysRevB.21.5312}.

\bibitem[{\citenamefont{Donley et~al.}(2002)\citenamefont{Donley, Claussen,
  Thompson, and Wieman}}]{Donley_2002}
\bibinfo{author}{\bibfnamefont{E.~A.} \bibnamefont{Donley}},
  \bibinfo{author}{\bibfnamefont{N.~R.} \bibnamefont{Claussen}},
  \bibinfo{author}{\bibfnamefont{S.~T.} \bibnamefont{Thompson}},
  \bibnamefont{and} \bibinfo{author}{\bibfnamefont{C.~E.}
  \bibnamefont{Wieman}}, \bibinfo{journal}{Nature}
  \textbf{\bibinfo{volume}{417}}, \bibinfo{pages}{529} (\bibinfo{year}{2002}),
  \urlprefix\url{https://doi.org/10.1038/417529a}.

\bibitem[{\citenamefont{Chin et~al.}(2010)\citenamefont{Chin, Grimm, Julienne,
  and Tiesinga}}]{Chin_2010}
\bibinfo{author}{\bibfnamefont{C.}~\bibnamefont{Chin}},
  \bibinfo{author}{\bibfnamefont{R.}~\bibnamefont{Grimm}},
  \bibinfo{author}{\bibfnamefont{P.}~\bibnamefont{Julienne}}, \bibnamefont{and}
  \bibinfo{author}{\bibfnamefont{E.}~\bibnamefont{Tiesinga}},
  \bibinfo{journal}{Rev. Mod. Phys.} \textbf{\bibinfo{volume}{82}},
  \bibinfo{pages}{1225} (\bibinfo{year}{2010}),
  \urlprefix\url{https://link.aps.org/doi/10.1103/RevModPhys.82.1225}.

\bibitem[{\citenamefont{Fernandes et~al.}(2010)\citenamefont{Fernandes,
  VanBebber, Bhattacharya, Chandra, Keppens, Mandrus, McGuire, Sales, Sefat,
  and Schmalian}}]{Fernandes_2010}
\bibinfo{author}{\bibfnamefont{R.~M.} \bibnamefont{Fernandes}},
  \bibinfo{author}{\bibfnamefont{L.~H.} \bibnamefont{VanBebber}},
  \bibinfo{author}{\bibfnamefont{S.}~\bibnamefont{Bhattacharya}},
  \bibinfo{author}{\bibfnamefont{P.}~\bibnamefont{Chandra}},
  \bibinfo{author}{\bibfnamefont{V.}~\bibnamefont{Keppens}},
  \bibinfo{author}{\bibfnamefont{D.}~\bibnamefont{Mandrus}},
  \bibinfo{author}{\bibfnamefont{M.~A.} \bibnamefont{McGuire}},
  \bibinfo{author}{\bibfnamefont{B.~C.} \bibnamefont{Sales}},
  \bibinfo{author}{\bibfnamefont{A.~S.} \bibnamefont{Sefat}}, \bibnamefont{and}
  \bibinfo{author}{\bibfnamefont{J.}~\bibnamefont{Schmalian}},
  \bibinfo{journal}{Phys. Rev. Lett.} \textbf{\bibinfo{volume}{105}},
  \bibinfo{pages}{157003} (\bibinfo{year}{2010}),
  \urlprefix\url{https://link.aps.org/doi/10.1103/PhysRevLett.105.157003}.

\bibitem[{\citenamefont{Fernandes et~al.}(2019)\citenamefont{Fernandes, Orth,
  and Schmalian}}]{Fernandes_2019}
\bibinfo{author}{\bibfnamefont{R.~M.} \bibnamefont{Fernandes}},
  \bibinfo{author}{\bibfnamefont{P.~P.} \bibnamefont{Orth}}, \bibnamefont{and}
  \bibinfo{author}{\bibfnamefont{J.}~\bibnamefont{Schmalian}},
  \bibinfo{journal}{Annual Review of Condensed Matter Physics}
  \textbf{\bibinfo{volume}{10}}, \bibinfo{pages}{133} (\bibinfo{year}{2019}),
  \eprint{https://doi.org/10.1146/annurev-conmatphys-031218-013200},
  \urlprefix\url{https://doi.org/10.1146/annurev-conmatphys-031218-013200}.

\bibitem[{\citenamefont{Fisher et~al.}(1973)\citenamefont{Fisher, Barber, and
  Jasnow}}]{Fisher_1973}
\bibinfo{author}{\bibfnamefont{M.~E.} \bibnamefont{Fisher}},
  \bibinfo{author}{\bibfnamefont{M.~N.} \bibnamefont{Barber}},
  \bibnamefont{and} \bibinfo{author}{\bibfnamefont{D.}~\bibnamefont{Jasnow}},
  \bibinfo{journal}{Phys. Rev. A} \textbf{\bibinfo{volume}{8}},
  \bibinfo{pages}{1111} (\bibinfo{year}{1973}),
  \urlprefix\url{https://link.aps.org/doi/10.1103/PhysRevA.8.1111}.

\bibitem[{\citenamefont{Nelson and Kosterlitz}(1977)}]{Nelson_1977}
\bibinfo{author}{\bibfnamefont{D.~R.} \bibnamefont{Nelson}} \bibnamefont{and}
  \bibinfo{author}{\bibfnamefont{J.~M.} \bibnamefont{Kosterlitz}},
  \bibinfo{journal}{Phys. Rev. Lett.} \textbf{\bibinfo{volume}{39}},
  \bibinfo{pages}{1201} (\bibinfo{year}{1977}),
  \urlprefix\url{https://link.aps.org/doi/10.1103/PhysRevLett.39.1201}.

\bibitem[{\citenamefont{Vanderstraeten
  et~al.}(2015)\citenamefont{Vanderstraeten, Mari\"en, Verstraete, and
  Haegeman}}]{Vanderstraeten_2015}
\bibinfo{author}{\bibfnamefont{L.}~\bibnamefont{Vanderstraeten}},
  \bibinfo{author}{\bibfnamefont{M.}~\bibnamefont{Mari\"en}},
  \bibinfo{author}{\bibfnamefont{F.}~\bibnamefont{Verstraete}},
  \bibnamefont{and} \bibinfo{author}{\bibfnamefont{J.}~\bibnamefont{Haegeman}},
  \bibinfo{journal}{Phys. Rev. B} \textbf{\bibinfo{volume}{92}},
  \bibinfo{pages}{201111} (\bibinfo{year}{2015}),
  \urlprefix\url{https://link.aps.org/doi/10.1103/PhysRevB.92.201111}.

\bibitem[{\citenamefont{Vanderstraeten
  et~al.}(2016)\citenamefont{Vanderstraeten, Haegeman, Corboz, and
  Verstraete}}]{Vanderstraeten_2016}
\bibinfo{author}{\bibfnamefont{L.}~\bibnamefont{Vanderstraeten}},
  \bibinfo{author}{\bibfnamefont{J.}~\bibnamefont{Haegeman}},
  \bibinfo{author}{\bibfnamefont{P.}~\bibnamefont{Corboz}}, \bibnamefont{and}
  \bibinfo{author}{\bibfnamefont{F.}~\bibnamefont{Verstraete}},
  \bibinfo{journal}{Phys. Rev. B} \textbf{\bibinfo{volume}{94}},
  \bibinfo{pages}{155123} (\bibinfo{year}{2016}),
  \urlprefix\url{https://link.aps.org/doi/10.1103/PhysRevB.94.155123}.

\end{thebibliography}
\end{document}